\documentclass[a4paper,11pt]{article}
\pdfoutput=1
\usepackage{jheppub}

\usepackage[T1]{fontenc}
\usepackage{amsfonts}
\usepackage{amsmath}
\usepackage{amssymb}
\usepackage{multirow}
\usepackage{graphicx}
\usepackage{subfig}
\usepackage{array,ragged2e}
\usepackage{scrextend}
\usepackage{tikz-cd}
\usepackage{mathtools}
\usepackage{xfrac} 

\usepackage{dsfont}
\usepackage{scrextend}
\usepackage{tikz-cd}
\usepackage{mathtools}

\DeclareMathOperator*{\IM}{Im}

\DeclareMathOperator{\Th}{Th}

\newcommand{\F}{{\mathbb{F}}}
\newcommand{\Z}{{\mathbb{Z}}}
\newcommand{\R}{{\mathbb{R}}}
\newcommand{\C}{{\mathbb{C}}}
\newcommand{\1}{{\mathds 1}}

\usepackage{xparse}

\newcommand{\vek}[1]{{\boldsymbol{#1}}}

\numberwithin{equation}{section}

\usepackage{graphicx}

\begin{document}

\title{Narain CFTs and Quantum Codes at Higher Genus}
\author[a]{Johan Henriksson,}
\emailAdd{johan.henriksson@df.unipi.it}
\affiliation[a]{Department of Physics, University of Pisa and INFN, \\Largo Pontecorvo 3, I-56127 Pisa, Italy}
\author[b]{Ashish Kakkar,}
\emailAdd{ashish.kakkar@uky.edu}
\affiliation[b]{Department of Physics and Astronomy, University of Kentucky,\\ Lexington, KY 40506, U.S.A}
\author[a]{Brian McPeak}
\emailAdd{brian.mcpeak@df.unipi.it}

\abstract{
  Code CFTs are 2d conformal field theories defined by error-correcting codes. Recently, Dymarsky and Shapere generalized the construction of code CFTs to include quantum error-correcting codes. In this paper, we explore this connection at higher genus. We prove that the higher-genus partition functions take the form of polynomials of higher-weight theta functions, and that the higher-genus modular group acts as simple linear transformations on these polynomials. We explain how to solve the modular constraints explicitly, which we do for genus 2. The result is that modular invariance at genus 1 \emph{and} genus 2 is much more constraining than genus 1 alone. This allows us to drastically reduce the space of possible code CFTs. We also consider a number of examples of ``isospectral theories'' -- CFTs with the same genus 1 partition function -- and we find that they have different genus 2 partition functions. 
  Finally, we make connection to some 2d CFTs known from the modular bootstrap. The $n = 4$ theory conjectured to have the largest possible gap, the $SO(8)$ WZW model, is a code CFT, allowing us to give an expression for its genus 2 partition function. We also find some other known CFTs which are not code theories but whose partition functions satisfy the same simple polynomial ansatz as the code theories. This leads us to speculate about the usefulness of the code polynomial form beyond the study of code CFTs.
}

 \maketitle
\flushbottom

\section{Introduction}

There is a fascinating relationship between error-correcting codes (ECCs) and two-dimensional conformal field theories. This relationship was first explored in the context of classical binary codes -- codes over $\mathbb F_2$ -- which are related to chiral meromorphic CFTs. Any doubly-even self-dual code defines an even self-dual lattice by ``Construction A'' of Leech and Sloane \cite{LeechSloane1971}, and such a lattice can be used to define a free meromorphic CFT by the work of Dolan, Goddard and Montague \cite{Dolan:1989kf,Dolan:1994st, Dolan1996}. In this relation, also reviewed in \cite{Dymarsky:2020qom}, a code of length $n$ provides an $n$-dimensional lattice, which gives rise to a meromorphic CFT of central charge $c=n$. The construction makes it extremely easy to compute the partition function of the resulting CFT. It takes the form
\begin{equation}
\label{eq:genus1formintro}
     Z^{(g=1)}(\tau)=\frac{W(\theta_3(q^2),\theta_2(q^2))}{\eta(\tau)^n} \, .
\end{equation}
Here $q=e^{2\pi i \tau}$ for the modular parameter $\tau$, $\eta(\tau)$ is the Dedekind eta function and $\theta_i(q^2)$ are Jacobi theta functions.
The function $W(x_0,x_1)$ is the so-called enumerator polynomial, a central object of this paper. It is trivially computable given the codewords defining the code. The conditions required for the partition function to be modular invariant (up to phases) follow immediately from simple properties of the code. These are the so-called MacWilliams identities \cite{Macwilliams1963}\footnote{Technically the MacWilliams identity of \cite{Macwilliams1963} refers to only the second transformation, while the first one follows from doubly-evenness. For convenience, we use the term ``MacWilliams identities'' to refer to both conditions.} 
\begin{align}
 W(x_0,x_1) = W(x_0,ix_1)\,,\qquad W(x_0,x_1)=W\left(\tfrac{x_0+x_1}{\sqrt2},\tfrac{x_0-x_1}{\sqrt2}\right)\, .
\end{align}
 The construction above is naturally extended to higher genus, where the generalized MacWilliams identities acting on the \emph{higher-weight enumerator polynomial} in $2^g$ variables guarantee genus~$g$ modular invariance of the corresponding partition function \cite{Henriksson:2021qkt}. The upshot is two-fold: firstly, since the enumerator polynomial is directly computable from the code, one gets immediate access to higher-genus partition functions of code CFTs; secondly, since the classification of the involved entities -- codes, lattices and CFTs -- is incomplete,\footnote{The codes involved in the construction -- binary double-even self-dual codes, also known as type~II codes -- have length $n$ divisible by 8 and have been classified for $n\leqslant40$ \cite{Betsumiya2012}. The corresponding lattices (even self-dual lattices) and CFTs (meromorphic CFTs) have been classified for $n=c\leqslant 24$ by \cite{Niemeier1973} and \cite{Schellekens:1992db} respectively.
} the enumerator polynomial form provides a simple way to analyze modular invariance as linear relations. 

Solving the constraints from higher-genus modular invariance amounts to characterizing the ring of invariant polynomials, which for genus $g\leqslant3$ determines the space of partition functions for meromorphic chiral CFTs via the theory of Siegel modular forms.\footnote{
For more details on the case at $g\geqslant4$ for chiral CFTs, see our previous work \cite{Henriksson:2021qkt}.
}  In \cite{Henriksson:2021qkt}, we showed that by imposing higher-genus modular invariance, the number of possible genus 1 partition functions of code CFT is greatly reduced. We also noted that there exist conformal field theories that do not derive from codes, but which still admit the ``code enumerator form'' of \eqref{eq:genus1formintro}, however with non-negative and non-integer coefficients. This holds for instance for all of the 71 meromorphic CFTs at $c=24$ classified by Schellekens \cite{Schellekens:1992db}.

In this paper, we turn to an interesting generalization of the above relation, namely between \emph{quantum error-correcting codes} and \emph{non-chiral CFTs}. This relation was spelled out in detail in \cite{Dymarsky:2020qom}. The starting point is a class of quantum error-correcting codes known as stabilizer codes. These have an equivalent description in terms of classical codes over the finite field $\F_4$. In general, one can consider codes defined over any finite field $F$, and it has been shown recently that ternary codes, defined over $F=\F_3$, define $\mathcal{N} = 1$ supersymmetric 2d CFTs \cite{Gaiotto:2018ypj}. Other work on the relation between error-correcting codes and conformal field theories include \cite{Dymarsky:2020bps, Gunaydin:2020ric,Dymarsky:2020pzc,Buican:2021uyp,Dymarsky:2021xfc,Furuta:2022ykh}. Interestingly, by the Gray map $\F_4$ can be related to $\F_2\oplus \F_2$, which is the first case of a series of constructions of Narain CFTs from codes over $\F_p\oplus \F_p$ for $p$ prime \cite{Yahagi:2022idq}.

Via the ``New Construction A,'' a code over $\F_4$ defines a Lorentzian lattice, on which a 2d non-chiral (``full'') CFT can be defined by the construction of Narain \cite{Narain:1985jj,Narain:1986am}. Compared to the chiral case discussed above, there are now several differences. In the New Construction A, the central charge $n = c=\bar c$ can take any positive integer value, and the phases showing up in the modular constraints always cancel, rendering completely modular invariant partition functions of the form
\begin{align}
\label{eq:Znonchiralintro}
Z_{\mathcal C}(q,\bar q)=\frac{W_{\mathcal C}(\theta_3(q)\theta_3(\bar q)+\theta_4(q)\theta_4(\bar q),\theta_3(q)\theta_3(\bar q)-\theta_4(q)\theta_4(\bar q), \theta_2(q)\theta_2(\bar q))}{2^n|\eta(\tau)|^{2n}}\,,
\end{align}
where $q=e^{2i\pi  \tau}$ and $ \bar q=e^{-2i\pi \bar \tau}$.
$W_{\mathcal C}(x_0,x_1,x_2)$ is called the refined enumerator polynomial, to be defined below.
On the other hand, this construction gives rise to a discrete set of CFTs, which lies inside but is not dense in the continuous Narain moduli space. For instance, in the simplest case of central charge $n=1$, the unique equivalence class of quantum codes gives rise to the Narain CFT of the compact boson at the radius $1$, whereas at $c=1$, Narain CFTs can be defined for any compactification radius. 

As in the chiral case, the partition function will be constructed by a sum over vectors in a lattice defined by the code, and it will be described in terms of higher-genus theta functions with known modular properties. The theory of such functions is less developed compared the chiral case, where indeed the theory of Siegel modular forms has led to strong constraints on meromorphic CFTs at low $c$ \cite{Gaberdiel:2009rd,Gaberdiel:2010jf,Keller:2017iql,Cho:2017fzo,Cardy:2017qhl}. 
Nevertheless, we shall see that the set of partition functions in the form dictated by \eqref{eq:Znonchiralintro} does capture some interesting theories also outside the class of code CFTs,\footnote{It remains possible that they may be code theories defined by some suitable generalization of New Construction A. See \cite{Buican:2021uyp} for a broader set of Narain CFTs with code counterparts. Throughout this paper, ``code theories'' means those defined by New Construction A.} to be discussed more in section~\ref{eq:extremalCFTs}, which may hint at the possibility to develop a theory of non-chiral modular forms.

The main novelty of this paper is to generalize the relation between quantum ECCs and full CFTs to higher-genus partition functions. This amounts to writing down higher-weight enumerator polynomials suitable for codes over general fields. In general, for a field $F$, there is a natural construction of a weight-$g$ enumerator polynomial in $|F|^g$ variables as a sum over $g$-tuples of codewords, see \eqref{eq:higherweightenumeratorpolygen} below. For the case at hand, where $F=\F_4$, the genus 1 version was called the complete enumerator polynomial in \cite{Dymarsky:2020qom}. However, we will see that the identical vanishing of some modular functions -- at genus 1 manifested by $\theta_1(q)=0$ -- suggests the use of a \emph{refined} enumerator polynomial in $2^{g-1}(2^g+1)=3,10,36,\ldots$ variables. (In fact, the counting is the same as the number of even spin structures in the Riemann surface, familiar from higher-loop superstring computations.) We will give a complete description of this construction, and give many details at genus 2.

\subsection*{A puzzle with fake partition functions}
One of our motivations to study the relation between quantum ECCs and full CFTs, was to resolve the following puzzle that emerged from the considerations at genus 1 in \cite{Dymarsky:2020qom}. At genus 1, the refined enumerator polynomial for a code $\mathcal C$ is homogeneous degree $n$, in three variables, and can be written as a sum over all codewords
\begin{align}
\label{eq:enumpolyintro}
    W_\mathcal{C}(x_0, x_1, x_2) \ = \ \sum_{c \in \mathcal{C}} x_0^{n - w_X(c) - w_Y(c) - w_Z(c)}  x_1^{w_Y(c)}x_2^{w_X(c) + w_Z(c)} \, ,
\end{align}
where we think of the codewords $c\in \mathcal C$ as vectors with entries in the set $\{\1,X,Y,Z\}$, and $w_t(c)$ counts the number of entries $c_i=t$.\footnote{As explained in detail in section~\ref{sec:review}, there is a direct relation between the elements $\{\1,X,Y,Z\}$, which represent Pauli matrices, and the elements $\{0,\omega, 1,\omega^2 \}$ that represent the finite field $\F_4$.} For codes to define CFTs, the enumerator polynomial defined in \eqref{eq:enumpolyintro} must satisfy the generalized MacWilliams identities \cite{Macwilliams1963,MacWilliams1977,Macwilliams1978, Shor1997, Nebe2006book}, which is to say that $W_\mathcal{C}(x_0, x_1, x_2) $ is invariant under
\begin{align}
\label{eq:MacWintro1}
    x_0 \mapsto \frac{1}{2}(x_0 + x_1 + 2x_2) \, , \quad x_1 \mapsto \frac{1}{2}(x_0 + x_1 - 2x_2) \, , \quad x_2 \mapsto \frac{1}{2} (x_0-x_1) \, ,
\end{align}
and 
\begin{align}x_1\mapsto -x_1.
\label{eq:MacWintro2}
\end{align}
The work of \cite{Dymarsky:2020qom} proposed to study solutions $W(x_0,x_1,x_2)$ of \eqref{eq:MacWintro1}--\eqref{eq:MacWintro2} without explicitly constructing any quantum code. Combining these equations with the assumption that $W(x_0,x_1,x_2)$ has non-negative integer coefficients, which must be the case for code CFTs, one finds a discrete set of solutions for each degree $n$. Only some of these solutions correspond to known codes. The simplest non-trivial example is at $n = 3$. In this case, \cite{Dymarsky:2020qom} found 10 solutions, however only four of them correspond to known codes:
\begin{align}
   \label{eq:intro code theories1}
     W_1^3 \ &= \ (x_0+x_2)^3 \, , \\ 
    W_1 W_2 \ & = \ (x_0+x_2)(x_0^2+x_1^2+2x_2^2) \, \\
    W_3 \ &= \ x_0^3+3 x_0 x_2^2+3 x_1^2 x_2+x_2^3\,, \\
    \tilde{W}_3 \ &= \ x_0^3+3 x_0 x_1^2+4 x_2^3\,.
    \label{eq:intro code theories4}
\end{align}
The six remaining polynomials also give rise to seemingly consistent partition functions with $n=3$, 
which we denote as ``fake partition functions.''\footnote{In fact, all of these fake partition functions can be constructed from linear combinations of the real theories. Specifically, all partition functions are non-negative linear combinations of the ones deriving from $W_1^3$. $W_3$ and $\tilde W_3$, see figure~\ref{fig:cequals3} below.}
These fake partition functions have enumerator polynomials in $x_i$ with non-negative integer coefficients and could therefore correspond to code enumerator polynomials. Furthermore, they have character decompositions with non-negative integer coefficients -- both in Virasoro characters and $U(1)^c\times U(1)^c$ characters. For example, the latter decomposition takes the form
\begin{equation}
    Z(\tau,\bar\tau)=\sum_{h,\bar h} d_{h,\bar h}\frac{q^{h}\bar q^{\bar h}}{\eta(\tau)^c\eta(\bar \tau)^c}\, .
\end{equation}
For instance, consider one of the fake theories,
\begin{align}
    W_{\mathrm{fake}} = \frac{2}{3} W_1^3 + \frac{1}{3} \tilde W_3 
\end{align}
The first few degeneracies of $U(1)^c\times U(1)^c$ primaries for $W_{\mathrm{fake}}$ read
\begin{align}
\begin{split}
    d_{0,0}&=1\,,\quad d_{\frac18,\frac18}=4 \,, \quad d_{\frac14,\frac14}=12\,, \quad d_{\frac34,\frac34}=8\, ,
    \\
    d_{1,1}&=12\,, \quad d_{1,0}=d_{0,1}=0\,, \quad \ldots\, .
    \end{split}
\end{align}
This fake theory, like all fake theories found by examining solutions to the Macwilliams identities, have all non-negative integer degeneracies, so we cannot prove that they do not correspond to CFTs by genus 1 considerations alone.
In this paper, we find that \emph{by considering the constraints from genus 2 modular invariance on code CFT partition functions, we can prove that a number of fake partition functions cannot be defined by error-correcting codes through New Construction A}. This includes all six fake theories at $n=3$. Continuing to higher values of central charge $n$, we find that the higher-genus constraints will not rule out all fake partition functions, but drastically reduce their number.

\subsection*{Structure of this paper}

The rest of this paper is structured as follows. In section~\ref{sec:review}, we review the construction of 2d CFTs from quantum error-correcting codes. This includes a number of elements. First we review classical codes over general fields, and how they can be used to define lattices and enumerator polynomials. Then we present the generalization to quantum codes. After giving a description of quantum error-correcting codes and an explanation of how error correction is achieved, we show how quantum ECCs define Lorentzian lattices and therefore Narain CFTs. A brief overview and a explicit example of the construction are given in section~\ref{sec:recap}.

Section~\ref{sec:Higher-genus} is where we explain how this construction may be extended to higher genus. This is essentially the main result of the paper. We show how higher-genus partition functions are related to the so-called higher-weight enumerator polynomials of the codes. These need to be evaluated with higher-weight Jacobi theta functions as arguments. We determine how modular invariance acts on these higher-weight enumerator polynomials. 

We use these observations to aid the classification of code CFTs. In section~\ref{sec:results}, we describe how to characterize the ring of polynomials defined by invariance under higher-genus modular transformations. This allows us to count the number of potentially valid genus 1 and genus 2 partition functions explicitly for $n \leqslant 6$. We show that only a fraction of the valid genus 1 partition functions arise as the factorization limit of valid genus 2 partition functions, which underlies our claim that genus 2 modular invariance is a strong constraint on the space of theories. We also discuss how we can resolve several sets of isospectral theories -- such theories have the same genus 1 partition function, but we find different genus 2 partition functions. Finally, in section~\ref{subsec:noncode} we point out the ``enumerator polynomial form'' is a useful ansatz for the partition function that applies to many non-code theories as well. We use it to construct modular invariant functions with a large gap in primaries.

\section{Quantum codes and Narain CFTs}
\label{sec:review}

The purpose of this section is to review the construction of CFTs from error-correcting codes. The original construction, due to \cite{Dolan:1989kf, Dolan:1994st}, associates chiral CFTs to binary codes. In this paper, we will be primarily interested in the more recent construction, due to \cite{Dymarsky:2020bps, Dymarsky:2020qom}, which associates non-chiral CFTs to classical codes over $\F_4$, or equivalently, quantum stabilizer codes. None of the material presented in this section is new: it is merely meant as a review of this work. A large part of the discussion also follows \cite{Nielsen} -- see that textbook for a more thorough introduction.

We have included a much briefer version of this discussion in section~\ref{sec:recap}. The reader who wants to proceed more quickly towards the results of this paper may prefer to start there.

\subsection{Classical codes over general fields}

Error-correcting codes are designed to encode information redundantly to protect against corruption. The classical example is the repetition code, where $0$ is encoded as $000$ and $1$ is encoded as $111$. In the information-theoretic context then, error-correcting codes should be thought of as a map from length-$k$ vectors to length-$n$ vectors, where $n>k$. However for our purposes, we shall simply think of them as a collection of codewords, \emph{i.e.} as the image of this map. For classical codes, these elements may be vectors over $\F_2$ (for the common case of binary codes) or any other finite field $F$. For quantum codes, the elements are spins, or qubits. Before turning to quantum codes, let us briefly review some facts about codes. These ideas are also reviewed in \cite{Elkies_lattices_i, Elkies_lattices_ii}, and more recently for a physics audience by \cite{Dymarsky:2020qom, Henriksson:2021qkt}.

Additive codes are those where the sum of any two codewords is a codeword. These may be easily specified by a generator matrix $G$:
\begin{align}
    c = G x, \qquad c \in \mathcal{C} \subset F^n,  \qquad x \in F^k \, .
\end{align}
Here we use $\mathcal{C}$ to denote the code and $c$ to denote its element codewords. It is clear then that $G$ must be an $n\times k$ matrix, and that there are $|F|^k$ codewords. Alternatively, a code may be specified by its parity check matrix $H$, defined to satisfy
\begin{align}
    H c = 0 \quad \text{if and only if } c \in \mathcal{C} \, .
\end{align}
The parity check matrix is an $(n-k) \times n$ matrix, and also satisfies $HG = 0$.  The parity check matrix is directly useful in error correction. If a codeword is corrupted, $c \to c' = c + e$, this can be easily detected by applying the parity check
\begin{align}
    Hc' = H(c + e) = H e \, .
\end{align}
We shall see that this step has a direct analogue when we discuss the case of quantum error-correcting codes.

The error-correction ability of a code is directly related to how far apart its codewords are, which is measured by the Hamming distance. The Hamming distance $d(c_1, c_2)$ for two codewords is defined as the $\ell^0$ norm, or the number of entries which are different. The Hamming distance for a code is defined as the minimum distance between any two codewords. An error-correcting code with length $n$, $ |F|^k$ elements, and Hamming distance $d$ is denoted as an $[n, k, d]$ code.

\paragraph{Enumerator polynomials} A coarse description of a code is provided by its enumerator polynomial, which counts the degeneracy of codewords. The most general such object is the \textit{complete enumerator polynomial}, defined by
\begin{align}
    W_\mathcal{C}(x_0, \ldots) = \sum_{c \in \mathcal C} \left( \prod_{i = 1}^n x_{c_i} \right) \, .
\end{align}
This is a function of $|F|$ variables, $x_0,\ldots,x_{|F|-1}$. The coefficient of each monomial $x_0^{n_0} x_1^{n_1},\ldots$ is the number of codewords with $n_0$ $0$s, $n_1$ copies of the first non-zero element of $F$, and so on. Here we imagine $x_0$ to correspond to $0\in F$, and the other $x_i$ to the other elements of $F$ in a fixed but otherwise arbitrary order.

For codes over $\F_2$, the enumerator polynomial simply counts the non-zero elements in each codeword. The number of non-zero elements is called the \textit{Hamming weight}, defined by $w(c) = d(c, 0)$. In this case, we can write the enumerator polynomial as
\begin{align}
    W_\mathcal{C}(x_0, x_1) = \sum_{c \in \mathcal{C}} x_0^{n - w(c)} x_1^{w(c)}.
    \label{eq:WchiralHamming}
\end{align}

Error-correcting codes can be used to define lattices by embedding them into a bigger vector space, and identifying lattice vectors as living in cosets defined by each codeword. The classic example of this is Construction A of Leech and Sloane \cite{LeechSloane1971}, which associates lattices in $\mathbb{R}^n$ to binary codes via
\begin{align}
    \Lambda(\mathcal{C}) = \left\{ \frac{v}{\sqrt{2}} \ \Big| \ v \in \mathbb{Z}^n, \ v \equiv c \text{ (mod 2) for some } c \in \mathcal C \right\}\, .
\end{align}
This identification leads to a relationship between the code enumerator polynomial and lattice theta function:\footnote{Recall the definition of the lattice theta function of a Euclidean lattice $\Lambda$: $\Theta_\Lambda(\tau)=\sum_{\lambda\in\Lambda} q^{\lambda^2/2}$.}
\begin{align}
    \Theta_\Lambda(\tau) \ = \ W_{\mathcal{C}}\left( \theta_3(q^2), \theta_2(q^2) \right) \, ,
\end{align}
\textit{i.e.} the lattice theta function can be found by substituting the polynomial variables $x_0, \ldots$ with Jacobi theta functions. This formula has generalizations for other fields, including the one relevant for our purposes: $\F_4$. It has elements $0, \, 1\,, \omega\, , \omega^2$ which satisfy
\begin{align}
    \omega*\omega^2 = 1, \qquad \omega+ 1 = \omega^2, \qquad 1 + 1 = \omega + \omega = \omega^2 + \omega^2 = 0\, .
\end{align}
We may think of $\omega$ as being a third-root of unity (in which case the last requirement, $x + x = 0$, must be input by hand). 

We shall primarily be interested in self-dual codes, since these will lead to modular-invariant partition functions. The dual of a code over a field $F$ is given by

\begin{align}
    C^\perp \equiv \{ a \in F^n \ | \ (a, c) = 0, \text{ for all} c \in \mathcal{C} \} \, .
\end{align}
Here the algebra is over $F$. For fields with characteristic 2, such as $\F_2$ or $\F_4$, this amounts to the requirement that $ (a, c) = 0 \text{ (mod 2)}$. The definition of duality depends on the definition of the inner product $(\, . \, )$ used. We will be interested in the inner-product defined by
\begin{align}
    a \cdot c = \sum_{i = 1}^n \bar a_i c_i + a_i \bar c_i \, ,
    \label{eq:dotproduct}
\end{align}
where $\bar a$ is the complex conjugate of $a$ when viewing $\omega$ as a third root of unity.

For a general field, code-duality acts on the Hamming enumerator polynomial by \cite{Macwilliams1963}
\begin{align}
    W_{\mathcal C^\perp}(x_0, x_1) = \frac{1}{|\mathcal{C}|} W_{\mathcal C}(x_0+ (|F|-1) x_1, \, x_0 - x_1) \, .
\end{align}
Self-dual codes are for which $\mathcal{C} = \mathcal{C}^\perp$. As a result, their enumerator polynomials are unchanged duality and $W_{\mathcal C}(x_0, x_1) = W_{\mathcal C^\perp}(x_0, x_1)$.

\paragraph{Example: extended Hamming [8,4,4] code}

The extended Hamming [8,4,4] code is a classical binary code defined by the generator matrix
\begin{align}
    G^T = \begin{pmatrix}
  0 & 1 & 0 & 1 & 0 & 1 & 0 & 1  \\
  0 & 0 & 1 & 1 & 0 & 0 & 1 & 1  \\
  0 & 0 & 0 & 0 & 1 & 1 & 1 & 1  \\
  1 & 1 & 1 & 1 & 0 & 0 & 0 & 0 
    \end{pmatrix}\, .
\end{align}
This code has $2^4 = 16$ codewords, which are the length $8$ vectors defined by multiplying all $16$ of the length $4$ binary vectors by $G$. The Hamming code is the unique doubly-even self-dual binary code of length $8$. Its enumerator polynomial is
\begin{align}
    W_{\text{Hamming}}(x_0, x_1) = x_0^8 + 14 x_0^4 x_1^4 + x_1^8 \, ,
\end{align}
were we used the formula \eqref{eq:WchiralHamming}.

\subsection{Quantum codes}

In direct analog to classical codes, quantum error-correcting codes are designed to protect quantum information, denoted by the state $ |\psi \rangle$, from corruption. Errors in quantum computation can take the form of an operator $E$ acting on the $ |\psi \rangle $. These errors are taken to be in the Pauli group $\mathcal{P}_n$ . The $n$-qubit Pauli group consists of tensor products of $I, X, Y \text{and}\ Z$ (these are the usual Pauli operators, also known by $I, \sigma_x, \sigma_y, \sigma_z$) and an overall phase of $\pm i \ \text{or} \ \pm 1$. Let $\mathcal{E}$ denote the linear space of errors acting on the Hilbert space. A subspace $C$ of the $n$-qubit Hilbert space is said to form a code iff
\begin{equation}
\langle\psi| E^\dagger E | \psi \rangle = c \left( E \right)     \, ,
\end{equation}
for all $E \in \mathcal{E} $ where $c \left( E \right)$ does not depend on the state $| \psi \rangle$. 
This is referred to as the Knill--Laflamme condition \cite{Knill_2000}.
\paragraph{Example: three qubit flip code}

Let us consider a very simplified example where the only possible errors are acting by $X$ which flips the states $|0 \rangle$ (spin up) and $|1 \rangle$ (spin down). We can protect against such errors by by encoding
\begin{align}
    a |0 \rangle + b |1 \rangle \quad \to \quad a |000 \rangle + b |111 \rangle \, .
\end{align}
The space of states spanned by $|000 \rangle$ and $|111 \rangle$ is called the code subspace, a subspace of the bigger three qubit Hilbert space. Now we transmit the message, and want to be able to correct any bit flips that may have happened. This process has two steps: 1) syndrome diagnosis, where we determine the errors, and 2) recovery, where we return the system to its initial state. 

Syndrome diagnosis is done via measurements of the four projection operators:
\begin{align}
\begin{split}
    P_0 \ &= \ |000\rangle\langle000| + |111\rangle\langle111| \, ,
    \\
    P_1 \ &= \ |100\rangle\langle100| + |011\rangle\langle011| \, , 
    \\
    P_2 \ &= \ |010\rangle\langle010| + |101\rangle\langle101| \, , 
    \\
    P_3 \ &= \ |001\rangle\langle001| + |110\rangle\langle110| \, . 
\end{split}
\end{align}

If no bit is flipped, then measuring $P_0$ will give 1. If only the $i^{\text{th}}$ bit is flipped, then only $P_i$ will give $1$. This code is constructed to detect and correct a maximum of 1 error on any qubit and will fail if there are two or three errors. Importantly, measuring these operators does not change the state (which is required for us to perform all four measurements). 

The second step is recovery. In this case it is very simple -- if a bit has been flipped, we apply $X$ to that bit to flip it back. 

These steps are directly analogous to the classical case: syndrome diagnosis for classical codes is performed by applying the parity check matrix $H$, and recovery is simply the step of interpreting the message as the closest codeword. This example was rather artificial, because we only allowed for particular kinds of errors, but it illustrates the error-detection and recovery steps. In particular, it shows that there are methods of detecting and correcting errors without destroying the state. Next we shall discuss a broader class of codes, which will be relevant to our interest in CFTs.

\subsubsection{Stabilizer codes}

A simple way to specify the code subspace is to specify a set of operators which stabilize that subspace. For example, the state
\begin{align}
    |\psi \rangle = \frac{1}{\sqrt{2}} \left( |00 \rangle + |11\rangle \right) \,
\end{align}
is stabilized by the operators $Z_1 Z_2$ and $X_1 X_2$, because these do not change $|\psi\rangle$. In fact, up to a phase, $|\psi\rangle$ is the unique such state. Therefore specifying the stabilizer, in this case, completely specifies the state $|\psi\rangle$. 

There are two basic requirements for a group to stabilize a non-trivial set of states: 1) that it does not include the element $-I$ (which obviously cannot stabilize any state) and 2) that all the elements commute. The latter follows because operators in $G_n$ either commute or anti-commute. If $g_1$ and $g_2$ anti-commute, then any element $|\psi \rangle$ which is stabilized by $g_1 g_2$ cannot be stabilized by $g_2 g_1$.

Now consider an operator $S$ which is in the stabilizer of the code subspace -- it has an eigenvalue $+1$ for any state $|\psi\rangle$. Then measuring $S$ will detect any errors $E$ which anticommute with $S$, because $S E |\psi\rangle = - E |\psi\rangle$. Each set of errors $E$ will anti-commute with a set of operators $S$, and this set of operators $S$ defines a stabilizer code. If $S$ is an Abelian subgroup of order $k$ of the Pauli group $\mathcal{P}_n$ and $-I\notin S$, then the space of states stabilized by all elements of $S$ is an $[[n,n-k,d]]$ quantum stabilizer code. Here, $d$ is the quantum Hamming distance of the code and is defined as the minimum weight of an operator which commutes with $S$ but is not in $S$. The weight of an operator is the number of $X$s, $Y$s, and $Z$s comprising it.

Stabilizer codes are related to classical codes over $\F_4$, discovered by Calderbank, Rains, Shor, and Sloane \cite{Calderbank:1996aj}. The key to this relation is the Gray map between $\F_4$ and $\F_2^2$, 
which associates
\begin{align}
\label{eq:Gray}
\begin{split}
    0 & \leftrightarrow (0, \, 0) \,,  \qquad 1 \ \leftrightarrow (1, \, 1) \, ,  \\
    \omega \ &\leftrightarrow (1, \, 0) \,,  \qquad \bar{\omega} \ \leftrightarrow (0, \, 1) \, .
    \end{split}
\end{align}
It is a $\F_2$-linear map that also relates the inner product \eqref{eq:dotproduct} to the symplectic inner product 
\begin{equation}\label{eq: def symplectic inner product  }
    (a,b)\cdot (a',b')=\sum_i a_ib'_i+a'_ib_i\,.
\end{equation}

This map can be used to relate classical codes over $\F_4$ with quantum stabilizer codes in the following way. Consider a classical code $\mathcal{C}$ over $\F_4$. Then $c \in \mathcal{C}$ is a length-$n$ vector whose entries are $0, 1, \omega,  \omega^2$. Through the Gray map, this can be related to a pair of vectors, $\alpha$ and $\beta$, with entries in $\F_2$. 
\begin{equation}
    c \,\leftrightarrow\, (\alpha,\beta)\,.
\end{equation}
The result is that each codeword $c \in \mathcal{C}$ can be used to specify a stabilizer in the related quantum code, $\mathcal{C}^*$ through the relation 
\begin{align}
    g = i^{\alpha \cdot \beta} \big( X_1^{\alpha_1} X_2^{\alpha_2}\ldots X_n^{\alpha_n} \big) \big( Z_1^{\beta_1} Z_2^{\beta_2}\ldots Z_n^{\beta_n} \big)\, .
    \label{eq:stabilizerfromvec}
\end{align}
We have specified each generator by the position of its $X$s and $Z$s, which are packaged into the binary vectors $\alpha$ and $\beta$. The relationship goes both ways and a set of stabilizers can be used to specify a set of codewords in $(\F_4)^n$, or equivalently elements of $(\F_2)^{2n}$.

Representing each generator as a pair of (row) vectors lets us represent the full set as the matrix $H = [\alpha | \beta ]$. This is broken into two $n \times (n-k)$ submatrices: A ``$1$'' in $\alpha_{ij}$ means that generator $g_i$ includes $X_j$. A ``$1$'' in $\beta_{ij}$ means that the generator $g_i$ includes $Z_j$. The presence of a $Y_j$ is indicated by a ``$1$'' in both $\alpha$ and $\beta$. 
\paragraph{Example: Steane [7,1,3] code} 

This is entirely specified by the stabilizer group generated by
\begin{align}
\begin{split}
    g_1 \ & = \ X_4 X_5 X_6 X_7 \, ,\\
    g_2 \ & = \ X_2 X_3 X_6 X_7 \, ,\\
    g_3 \ & = \ X_1 X_3 X_5 X_7 \, ,\\
    g_4 \ & = \ Z_4 Z_5 Z_6 Z_7 \, ,\\
    g_5 \ & = \ Z_2 Z_3 Z_6 Z_7 \, ,\\
    g_6 \ & = \ Z_1 Z_3 Z_5 Z_7\, .
\end{split}
\end{align}
As an example, the parity check matrix of the Steane code is
\begin{align}
\label{eq:Steane}
    H = \left( \begin{array}{@{}c|c@{}}
   \begin{matrix}
      0 & 0 & 0 & 1 & 1 & 1 & 1 \\
      0 & 1 & 1 & 0 & 0 & 1 & 1 \\
      1 & 0 & 1 & 0 & 1 & 0 & 1 \\
      0 & 0 & 0 & 0 & 0 & 0 & 0 \\
      0 & 0 & 0 & 0 & 0 & 0 & 0 \\
      0 & 0 & 0 & 0 & 0 & 0 & 0      
   \end{matrix} 
      &
    \begin{matrix}
      0 & 0 & 0 & 0 & 0 & 0 & 0 \\
      0 & 0 & 0 & 0 & 0 & 0 & 0 \\
      0 & 0 & 0 & 0 & 0 & 0 & 0\\
      0 & 0 & 0 & 1 & 1 & 1 & 1 \\
      0 & 1 & 1 & 0 & 0 & 1 & 1 \\
      1 & 0 & 1 & 0 & 1 & 0 & 1
   \end{matrix} 
\end{array} \right)\, .
\end{align}
Clearly $H$ is the quantum version of the classical parity check matrix. The condition that  the stabilizers form an Abelian subgroup  equivalent to 
\begin{equation}
    H g H^\text{T} = 0\, ,
\end{equation}
where 
\begin{equation}
    g= \begin{pmatrix}
    0 & I_{\text{n}\times\text{n}}  \\
    I_{\text{n}\times\text{n}} & 0
    \end{pmatrix}\, .
\end{equation}
%
This equation is equivalent to 
\begin{align}
    g_i g_j - g_j g_i = 0 \quad & \Leftrightarrow  \quad \alpha_i \cdot \beta_j - \alpha_j \cdot \beta_i \equiv 0 \text{ (mod 2) } \\
    \quad &\Leftrightarrow \quad  \bar{c}_i \cdot c_j - c_i \cdot \bar{c}_j \equiv 0 \text{ (over $\mathbb{F}_4$) } \, .
\end{align}

The second line here amounts to the requirement that the code over $\mathbb{F}_4$ is self-orthogonal, meaning $\mathcal C \subset \mathcal{C}^{\perp}$. Self-duality ($\mathcal{C} = \mathcal{C}^\perp$) is a stronger requirement. A vector in $\mathbb{F}_4^n$ can be thought of as having real dimension $2n$ through the Gray map. Therefore if $\mathcal{C}$ is an $[n, m, d]$ code, then its dual will be an $[n, 2n-m, d]$ code. As a result, self-dual codes over $\F_4$ must be $[n, n, d]$ codes. In fact, classical $[n, m, d]$ codes define quantum $[[n, n-m, \tilde d]]$ codes, so we see that self-dual codes over $\F_4$ define quantum codes which cannot actual transmit any information. \footnote{These are sometimes referred to as error-detection, rather than error-correction, protocols.} From here on out, we will consider $k$ to define the size of the quantum $[[n, k, \tilde d]]$ code. For self-dual codes, $k = 0$.

The codewords are the vectors stabilized by the generators defined by $H$, and the space of all linear combinations of codewords is called the ``code subspace.'' Just as in the classical case, the codewords are given by the kernel of $H$, with multiplication defined in \eqref{eq: def symplectic inner product  }. 

To make this more precise, we can define a binary ``generator matrix'' $G$ of dimension $2n \times( n+k)$ whose columns form a basis of codewords. It must be defined to satisfy
\begin{equation}
    HgG=0\,,
\end{equation}
for all $g$. $G$ can be chosen so that its first $n-k$ rows coincide with those of $H$. The remaining rows, spanning logical operations on the code subspace, will not matter for our purposes because we have $k = 0$. 

For the Steane code, the logical operators (operators which commute with the stabilizers but are not in the code subspace) are 
\begin{align}
\begin{split}
    X_L \ & = \ X_1 X_2 X_3 X_4 X_5 X_6 X_7 \, ,\\
    Z_L \ & = \ Z_1 Z_2 Z_3 Z_4 Z_5 Z_6 Z_7 \, .
    \end{split}
\end{align}
The logical states or eigenstates of these logical operators are 
\begin{align}
\begin{split}
    & |0 \rangle_{\text{L}} \quad = \quad \frac{1}{\sqrt{8}} \Big( |0000000\rangle + |1010101\rangle + |0110011\rangle + |1100110\rangle \\
    & \qquad \qquad \qquad + |0001111\rangle + |1011010\rangle + |0111100\rangle + |1101001\rangle \Big)\, , \\
    & |1 \rangle_{\text{L}} \quad = \quad \frac{1}{\sqrt{8}} \Big( |1111111\rangle + |0101010\rangle + |1001100\rangle + |0011001\rangle \\
    & \qquad \qquad \qquad + |1110000\rangle + |0100101\rangle + |1000011\rangle + |0010110\rangle \Big) \, .
    \label{eq:Steane_encoding}
\end{split}
\end{align}

The Steane code is an example of Calderbank--Shor--Steane (CSS) codes.  These codes will be interesting to us because they may be constructed from classical codes. In particular, consider two classical binary codes, an $[n, k_1]$ code $\mathcal{C}_1$, and an $[n, k_2]$ code $\mathcal{C}_2$, and which satisfy $\mathcal{C}_2 \subseteq \mathcal{C}_1$. Then we can form an $[n, k_1 - k_2]$ code, denoted $CSS(\mathcal{C}_1, \mathcal{C}_2)$ in the following way: For a given codeword of $x \in \mathcal{C}_1$, we define
\begin{align}
    |x + \mathcal{C}_2 \rangle = \frac{1}{\sqrt{|\mathcal{C}_2|}} \sum_{y \in \mathcal{C}_2} |x + y \rangle \, .
\end{align}
If we do this for each codeword, we will end up with $k_1$ codewords in the quantum code. But many of these may be the same -- this will happen for two codewords $x$ and $x' \in \mathcal{C}_1$ whenever $x - x' \in \mathcal{C}_2$. So in fact, the code $\mathcal{C}_1$ breaks into cosets determined by the structure of $\mathcal{C}_2$, and the resulting code has $2^{k_1 - k_2}$ unique codewords.

The Steane code is able to correct ``arbitrary single-qubit errors.'' This includes a phase flip (applying $Z$ to a single bit) and a single bit flip (applying $X$ to a single bit). We will define $e_1$ to be a vector with a single 1, which denotes the position of the phase flip, and a similar vector $e_2$ to denote the position of the bit flip. 

If the codewords becomes corrupted, then $|\psi \rangle \to |\psi' \rangle$ becomes 
\begin{align}
    \frac{1}{\sqrt{|\mathcal{C}_2|}} \sum_{y \in \mathcal{C}_2} |x + y \rangle \to \frac{1}{\sqrt{|\mathcal{C}_2|}} \sum_{y \in \mathcal{C}_2} (-1)^{(x+y)\cdot e_1} |x + y + e_2\rangle \, .
\end{align}

Syndrome diagnosis for the bit flip is accomplished by first adding auxiliary qubits to the system, i.e. to write $|x + y + e_2\rangle $ as $ |x + y + e_2\rangle |0 \rangle_{\mathrm{aux}}$ and then mapping\footnote{It is a non-trivial fact that the transformation $|x \rangle |0 \rangle_{\mathrm{aux}} \to |x \rangle |Hx \rangle_{\mathrm{aux}}$ can always be accomplished using a quantum circuit composed of CNOT gates.} $|x + y + e_2\rangle |0 \rangle_{\mathrm{aux}} \to |x + y + e_2\rangle |H (x + y + e_2) \rangle_{\mathrm{aux}} = |x + y + e_2\rangle |H e_2 \rangle_{\mathrm{aux}}$.

Detecting the phase flip errors is almost identical, after using a trick: a phase flip, which acts by $|0\rangle \to |0\rangle$, $|1 \rangle \to -|1\rangle$, acts the same as a bit flip in the basis $|+\rangle = |0\rangle + |1 \rangle$, $|-\rangle = |0\rangle - |1 \rangle$. So we change basis (formally, apply a Hadamard gate) and then we see that we can detect this error using the same procedure as for the bit flip errors.



%
%



\subsubsection{Enumerator polynomials for quantum codes}

Using the Gray map, we can define various types of enumerator polynomials for quantum codes. For classical codes, the enumerator polynomial will count the number of $0$s, $1$s, $\omega$s, and $ \omega^2$s, of each codeword. For the quantum code, this corresponds to counting the number of $I$s, $Y$s, $X$s, and $Z$s in each stabilizer. Therefore we define the weights
\begin{align}
    w_x(c) = \vec 1 \cdot \alpha, \quad w_y(c) = \alpha \cdot \beta, \quad w_z(c) =\vec 1 \cdot \beta \, .
\end{align}
As a result, the complete enumerator polynomial can be written
\begin{align}
\label{eq:genuPcomplete}
    W_\mathcal{C}(x_0,x_1,x_2,x_3) \ = \ \sum_{c \in \mathcal{C}} x_0^{n - w_x(c) - w_y(c) - w_z(c)} x_1^{w_x(c)} x_2^{w_y(c)} x_3^{w_z(c)}\, .
\end{align}
However we will not need this. By an argument from \cite{Dymarsky:2020qom}, which we will revisit in section~\ref{sec:Higher-genus}, it is convenient to instead study the 
refined enumerator polynomial, 
\begin{align}
    W_\mathcal{C}(x_0, x_1, x_2) \ = \ \sum_{c \in \mathcal{C}} x_0^{n - w_x(c) - w_y(c) - w_z(c)} x_1^{w_y(c)} x_2^{w_x(c) + w_z(c)} \, .
    \label{eq:defREP}
\end{align}
Self-duality, on the level of the refined enumerator polynomial, takes the form of the requirement that $W_\mathcal{C}(x_0, x_1, x_2)$ is invariant under\footnote{This follows from a general identity for the complete enumerator polynomial \cite{MacWilliams1977} (Theorem~10 of chapter~5), see also \cite{Macwilliams1978}, Theorem~8.}
\begin{align}
\label{eq:MacWilliamsG1}
    x_0 \to \frac{1}{2}(x_0 + x_1 + 2x_2) \, , \quad x_1 \to \frac{1}{2}(x_0 + x_1 - 2x_2) \, , \quad x_2 \to \frac{1}{2} (x_0-x_1) \, .
\end{align}

We need another criterion beyond self-duality: a stabilizer code is called \textit{real} if the stabilizers are all real, \textit{i.e.} the number of $Y$s in each stabilizer is even. This requires $w_y(c)$ to be an even number, which means that real codes are invariant under 
\begin{align}
\label{eq:MacWilliamsExtra}
    x_1 \to - x_1 \, .
\end{align}

\subsection{CFTs from quantum codes}

Having reviewed most of the required elements, we are now ready to explain the so-called New Construction A, due to Dymarsky and Shapere, which is a construction of non-chiral, ``full,'' CFTs from stabilizer codes, or equivalently from codes over $\F_4$. The central point is that each code defines a lattice through
\begin{align}
\label{eq:Lorentzianlatticefromcode}
    \Lambda(\mathcal C) = \left\{ \frac{v}{\sqrt 2}  \ \Big| v \in \mathbb{Z}^{2n}, \, v \equiv (\alpha, \beta) \text{ (mod 2) for some } (\alpha, \beta) = c \in \mathcal{C} \right\} \, .
\end{align}
To define a CFT, we would like to think of this as a Lorentzian lattice. 
This can be done by embedding $\Lambda(\mathcal C)$ in $\R^{2n}$. If we use $v \leftrightarrow (a, b)$, which is the same as the $c\leftrightarrow (\alpha, \beta)$ basis for the codewords, then we use the symplectic metric
\begin{align}
    g = \begin{pmatrix} 0 & I \\ I & 0 \end{pmatrix} \, ,
\end{align}
leading to $|v|^2 = 2 a \cdot b $. This can be transformed to coordinates where we have the usual Lorentzian metric, $|v|^2 = p_L^2 - p_R^2$, by defining
\begin{align}
 \label{eq:pLpR}
    p_L = \frac{a + b}{\sqrt{2}}, \qquad p_R = \frac{a - b}{\sqrt{2}} \, .
\end{align}
When considered as a Lorentzian lattice, the following result \cite{Dymarsky:2020qom} follows: 
\begin{itemize}
    \item The lattice defined by a code will be self-dual if and only if the code is self-dual.
    \item A lattice defined by a code will be even if and only if the code is real.
\end{itemize}

Lattices may be characterized by their theta functions, which for a Lorentzian lattice takes the form\footnote{Here we define the theta function the following way. For a Lorentzian lattice in $(n,n)$ signature, move to coordinates where the metric is of the form $g=\mathbb I_{n,\times n}\oplus (-\mathbb I_{n\times,n})$, so that any lattice vector can be written $v=(\ell,r)$. Then the lattice theta function is defined as $\Theta_\Lambda=\sum_{v\in\Lambda}q^{\ell^2/2}\bar q^{r^2/2}$.}
\begin{align}
    \Theta_{\Lambda}(\tau, \bar \tau)\  = \ \sum_{v \in \Lambda} q^{p_L^2 / 2}\,  \bar{q}^{p_R^2 / 2} \, , \qquad q = e^{2 \pi i \tau}\, , \quad \bar{q} = e^{-2 \pi i \bar \tau} \, .
    \label{eq:thetafunction}
\end{align}
Just as in the case of Euclidean lattices, the theta function for a Lorentzian lattice is related to the enumerator polynomial of its defining code. In this case, the relationship takes the form \cite{Dymarsky:2020qom}:
\begin{align}
    \Theta_{\Lambda(\mathcal C)}(\tau, \bar \tau) = W_{\mathcal C}(\Theta_3+\Theta_4,\Theta_3-\Theta_4,\Theta_2)
    \label{eq:EPtoTheta}\, ,
\end{align}
where we have defined
$\Theta_m(\tau,\bar\tau)=\theta_m(e^{2\pi i\tau})\theta_m(e^{-2\pi i \bar\tau})$. We will give the derivation of this formula in section~\ref{subsec:derivation} as a special case of the general-genus result.

\subsubsection{Narain CFTs}

Consider now the theory of $n$ free bosons compactified on a lattice $\Gamma$ \cite{Narain:1985jj,Narain:1986am}, \textit{i.e.} moving freely in $\mathbb{R}^n/\Gamma$. This theory is described by the action
\begin{align}
    S = - \frac{1}{4 \pi \alpha'} \int dt d \sigma \sqrt{-g} \left(\partial_{\mu} \Phi^I \partial^\mu \Phi^I + \epsilon^{\mu \nu} B_{IJ} \partial_{\mu} \Phi^I \partial_\nu \Phi^J \right)\, .
\end{align}
The antisymmetric field $B$ is required to construct the most general theory of this type.  

Consider now the case where spacetime is also a 2d torus. Then periodicity requires that $\vec \Phi(t, \sigma) \sim  \vec \Phi(t, \sigma + 2 \pi)$. But the lattice compactification implies we have also identified $ \vec \Phi(t, \sigma) \sim \vec \Phi(t, \sigma) + 2 \pi \vec \lambda$, where $\vec \lambda \in \Gamma$ (note that $\Gamma$ is different from $\Lambda$, which is formed from $\Gamma$ and $\Gamma^*)$. So the most general possibility is 
\begin{align}
    \vec \Phi(t, \sigma + 2 \pi) = \vec\Phi(t, \sigma) + 2 \pi \vec \lambda \, ,
    \label{eq:periodicity}
\end{align}
where $\vec \lambda$ is zero or any other element of $\Gamma$. Now consider the following solution to the equations of motion:
\begin{align}
    \vec \Phi_L(t + \sigma) = \frac{1}{2} \vec \Phi(0,0) + \frac{1}{2}\alpha' \vec p_L(t + \sigma) + \frac{1}{2} i \sum_{n \neq 0} \frac{\vec a_n}{n}e^{-in(t + \sigma)} \, ,
    \\
    \vec \Phi_R(t + \sigma) = \frac{1}{2} \vec \Phi(0,0) + \frac{1}{2}\alpha' \vec p_R(t - \sigma) + \frac{1}{2} i \sum_{n \neq 0} \frac{\vec b_n}{n}e^{-in(t + \sigma)}\, ,
    \label{eq:stringsol}
\end{align}
where $\vec \Phi(t, \sigma) = \vec \Phi_L(t + \sigma) + \vec \Phi_R(t - \sigma)$. From the solution, we see that if $\sigma \to \sigma + 2 \pi$, then 
\begin{align}
    \vec \Phi \to \vec \Phi + \pi \alpha' (\vec p_L - \vec p_R) \, .
\end{align}
Thus if the periodicity condition of~\eqref{eq:periodicity} is to be satisfied, we must have
\begin{align}
    \frac{1}{2} \alpha' (\vec p_L - \vec p_R) =  \vec \lambda \in \Gamma \,.
\end{align}
Furthermore, compactification on the lattice $\Gamma$ implies that the momenta $\vec P$ is in the dual lattice $\Gamma^*$.
By computing the canonical momentum, we find that
\begin{align}
    \vec V = \alpha' \vec P + B \vec \lambda \, .
\end{align}
$\vec V$ is defined as the coefficient multiplying $t$ in the solution $\vec \Phi(t, \sigma)$. From~\eqref{eq:stringsol}, it must be
\begin{align}
 \vec V = \frac{1}{2} \alpha' (\vec p_L + \vec p_R) \, .
\end{align}
Solving for $\vec p_L$ and $\vec p_R$, we simply find
\begin{align}
    \vec p_L = \vec P + \frac{1}{\alpha'} (B + I) \vec \lambda, \qquad \vec p_R = \vec P + \frac{1}{\alpha'} (B - I) \vec \lambda \, .
\end{align}
The set of all $(\vec p_L,\vec p_L)$ in this parametrization forms the lattice $\Lambda$.
From here on, we will set $\alpha' = 2$ to keep $\vec p_L$ and $\vec p_R$ dimensionless, as they are in the previous subsection.

The Narain theories have a $U(1)^n \times U(1)^n $ symmetry, corresponding to moving $\vec \Phi_L$ or $\vec \Phi_R$ around the compact directions. The primary operators with respect to this symmetry are 
\begin{align}
    V_{p_L, p_R} = e^{i \vec p_L \vec \Phi_L} e^{i \vec p_R \vec \Phi_R} \, .
    \label{eq:vertex}
\end{align}
Since the elements of the $v \in \Lambda$ lattice are labeled by $\vec p_L$ and $\vec p_R$, we see that we have a single primary for each lattice vector. The weights of these primaries are simply $h = p_L^2 / 2$, $\bar h = p_R^2 / 2$.

The characters of the $U(1)^n \times U(1)^n $ symmetry group are
\begin{align}
\label{eq:characters}
    \chi_{h,\bar h}(\tau,\bar \tau) = \frac{q^h\bar q^{\bar h}}{\eta(\tau)^n\eta(\bar\tau)^n} \, .
\end{align}
The final result is the partition function, which equals
\begin{align}
    Z(\tau, \bar \tau) \ = \ \sum_{ v\in \Lambda} \frac{q^{p_L^2/2} \bar{q}^{p_R^2/2}}{\eta(\tau)^n \eta(\bar \tau)^n} \, .
\end{align}
The numerator of this sum is precisely the lattice theta function introduced in~\eqref{eq:thetafunction}. Combining this with the result~\eqref{eq:EPtoTheta} yields a formula for the genus 1 partition function:
\begin{align}
Z_{\mathcal C}=\frac{W_{\mathcal C}(\Theta_3+\Theta_4,\Theta_3-\Theta_4, \Theta_2)}{2^n|\eta(\tau)|^{2n}}\, ,
\end{align}
for the CFT defined by the code. The numerator in this expression is simply the refined enumerator polynomial, evaluated at combinations of the Jacobi theta functions in the notation $\Theta_m(\tau,\bar\tau)=\theta_m(e^{2\pi i\tau})\theta_m(e^{-2\pi i \bar\tau})$. 

\subsubsection{Code theories}
\label{subsec:CodeTheories}

We have now reviewed how a code $\mathcal{C}$ defines a Lorentzian lattice $\Lambda(\mathcal{C})$, via~\eqref{eq:Lorentzianlatticefromcode}, and how a lattice $\Lambda$ defines a CFT, essentially through the definition of the vertex operators ~\eqref{eq:vertex}, where $(\vec p_L, \vec p_R) \in \Lambda$. Thus it is clear how to associate a CFT to a code. 

In practice, however, different codes may define the same theory, and it may be useful to have a way of classifying all distinct theories. There is a large group of code equivalences, which are transformations between codes which define the same theory. They include permuting the components of codewords and swapping $X_i$ and $Z_i$ for any $i$. These lead to different lattices which are related by T-dualities, so the corresponding CFT will ultimately be the same. In fact, all T-dualities which relate two code theories are of this form, \textit{i.e.} permutations of components and swaps of $X$ and $Z$ \cite{Dymarsky:2020qom}.

Recall now that self-dual real codes can be specified by $n$ pairs of (row) vectors $(\alpha_i, \beta_i)$, which define the code's generator matrix:
\begin{align}
    G^T = \left( \begin{array}{@{}c|c@{}}
   \begin{matrix}
      \alpha_1 \\
      \alpha_2 \\
      \vdots \\
      \alpha_n
   \end{matrix} 
      &
    \begin{matrix}
      \beta_1 \\
      \beta_2 \\
      \vdots \\
      \beta_n
   \end{matrix} 
\end{array} \right)\, .
\end{align}
One of the main results of \cite{Dymarsky:2020qom} is that, due to code equivalences, every code theory  defined from New Construction A can be described by a code whose generator matrix has the form
\begin{align}
    G^T = \left( \begin{array}{@{}c|c@{}}
   \begin{matrix}
      B
   \end{matrix} 
      &
    \begin{matrix}
      I
   \end{matrix} 
\end{array} \right)\, ,
\end{align}
where $B$ is an antisymmetric binary matrix. Codes in this form are called $B$-form codes. Therefore the result can be stated in the following way: any real self-dual code is equivalent to a $B$-form code. The generators for a $B$-form code take the form
\begin{align}
    g_i = Z_i \prod_{j = 1}^n (X
    _j)^{B_{ij}}\, .
\end{align}

One way to organize the set of possible matrices $B$ is through graphs. A graph can be defined by an \textit{adjacency matrix} $M$ where the entry $M_{ij}$ contains information about the link between node $i$ to node $j$. $B$ is a binary antisymmetric matrix, which is equivalent (mod 2) to a symmetric matrix with zeroes on the diagonal. Hence the resulting graph is undirected and has no self-links. The end result is that, due to code equivalences, each code theory can be represented by a binary $n\times n$ matrix $B$ or by an undirected graph with $n$ nodes. We shall use both throughout this paper.\footnote{The $B$-form, and related graph, are not necessarily unique. This is due to the existence of certain T-dualities that relate two different $B$-form codes. These act on the graphs. This issue is important for the classification of code theories in \cite{Dymarsky:2020qom}.}

\subsection{Recap and overview}
\label{sec:recap}

Since this section includes a number of diverse elements, we will conclude with an executive summary.

\paragraph{Codes} A stabilizer code is defined by its codewords, which take the form 
\begin{align}
    g = i^{\alpha \cdot \beta} \big( X_1^{\alpha_1} X_2^{\alpha_2}\ldots X_n^{\alpha_n} \big) \big( Z_1^{\beta_1} Z_2^{\beta_2}\ldots Z_n^{\beta_n} \big)\, .
    \label{eq:grecap}
\end{align}
Here $n$ is the dimension of the code space. The number of generators is $k$, so there are $2^k$ codewords. The set of codewords can then be specified by the length-$n$ binary vectors $\alpha$ and $\beta$. There is one pair $(\alpha, \beta)$ for each codeword.

The Gray map~\eqref{eq:Gray} associates tuples in $\F_2$ with elements of $\F_4$. This allows a pair $(\alpha, \beta)$ of binary vectors to be combined into a vector over $\F_4$, which we denote $c$. We then think of the collection of $c$s as elements of a code over $\F_4$. 

\paragraph{Enumerator polynomials} There is a natural definition of enumerator polynomials for codes over general fields, so the Gray map allows us to define an enumerator polynomial for our quantum codes. First we define the weights
\begin{align}
    w_x(c) = \vec 1 \cdot \alpha, \quad w_y(c) = \alpha \cdot \beta, \quad w_z(c) = \vec 1 \cdot \beta \, .
\end{align}
The enumerator polynomial of interest for us is the refined enumerator polynomial 
\begin{align}
    W_\mathcal{C}(x_0, x_1, x_2) \ = \ \sum_{c \in \mathcal{C}} x_0^{n - w_x(c) - w_y(c) - w_z(c)} x_1^{w_y(c)} x_2^{w_x(c) + w_z(c)} \, .
    \label{eq:recapdefREP}
\end{align}
The abelian group structure of the stabilizer codes already ensures that all $\F_4$ codes defined this way are additive and self-orthogonal ($\mathcal{C} \subset \mathcal{C}^\perp$). The $\F_4$ code will be even provided that the quantum codes are real, meaning all generators $g$ in the form~\eqref{eq:grecap} are real. Self-duality is equivalent to the requirement that the code is self-orthogonal, which is automatically satisfied, plus the requirement that $|\mathcal{C}| = |\mathcal{C}^\perp|$. This means that the $\F_4$ code is an $[n, n, d]$ code, so the quantum code has $n$ generators. The enumerator polynomials of such codes are invariant under the transformations
\begin{align}
    x_0 \to \frac{1}{2}(x_0 + x_1 + 2x_2) \, , \quad x_1 &\to \frac{1}{2}(x_0 + x_1 - 2x_2) \, , \quad x_2 \to \frac{1}{2} (x_0-x_1) \, ,
    \label{eq:recapMac1}
\end{align}
and
\begin{align} 
    x_1 &\to - x_1 \, .
    \label{eq:recapMac2}
\end{align}

\paragraph{Lattices} 

Codes over $\F_4$ define lattices via
\begin{align}
    \Lambda(\mathcal C) = \left\{ \frac{v}{\sqrt 2}  \ \Big| v \in \mathbb{Z}^{2n}, \, v \equiv (\alpha, \beta) \text{ (mod 2) for some } (\alpha, \beta) = c \in \mathcal{C} \right\} \, .
\end{align}
It follows from the definition that the lattice will be self-dual if the code is self-dual (with respect to the symplectic metric), and it will be even if the code is real.

Lattices are characterized by their theta functions, defined in \eqref{eq:thetafunction}. The lattice theta function can be directly computed from the error-correcting code via the formula
\begin{align}
    \Theta_{\Lambda(\mathcal C)}(\tau, \bar \tau) = W_{\mathcal C}(\Theta_3+\Theta_4,\Theta_3-\Theta_4, \Theta_2)\, ,
\end{align}
with $\Theta_m(\tau,\bar\tau)=\theta_m(e^{2\pi i\tau})\theta_m(e^{-2\pi i \bar\tau})$.

\paragraph{CFT partition function}

The code CFT is defined as the Narain theory associated to the lattice, 
\begin{equation}
\label{eq:PFg1char}
    Z_{\mathrm{CFT}}=\sum_{h,\bar h} \chi_{h,\bar h}\, ,
\end{equation}
for the $U(1)^n\times U(1)^n$ characters \eqref{eq:characters}.
Via the lattice construction, we related this to the enumerator polynomial, and arrived at the expression
\begin{equation}
\label{eq:PFg1}
    Z_{\mathcal C}=\frac{W_{\mathcal C}(\Theta_3+\Theta_4,\Theta_3-\Theta_4,\Theta_2)}{2^c|\eta(\tau)|^{2c}}\, ,
\end{equation}
for the CFT defined by the code. The numerator in this expression is simply given by the refined enumerator polynomial; the denominator is a universal factor, corresponding to a one-loop determinant.

The conformal weights in \eqref{eq:PFg1char} are given by 
\begin{equation}
    (h,\bar h)=\left(\frac12\vec p_L^{\,2}, \frac12\vec p_R^{\,2}\right), \quad (\vec p_L,\vec p_R)=\left(\frac{a+b}2, \frac{a-b}2\right), \quad a=2 \vec P+B\vec \lambda, \quad b=\vec \lambda\, ,
\end{equation}
where $\vec \lambda$ and $\vec P$ take all values in the integer lattice $\mathbb Z^n$. A parametrization in terms of the codewords $c\leftrightarrow (\alpha,\beta)$ (using the Gray map), is
\begin{equation}
    \qquad (\vec p_L,\vec p_R) = \left(a+\frac\alpha2,b+\frac\beta2 \right), \qquad \quad a,b\in \mathbb Z^n, \quad  (\alpha,\beta)\leftrightarrow c\, .
\end{equation}
The sum over all $(\vec p_L,\vec p_R)$ then becomes a sum over all codewords $c$ and all integer values for $a$ and $b$.

Modular invariance of the partition function \eqref{eq:PFg1} follows from the MacWilliams identities ~\eqref{eq:recapMac1} and~\eqref{eq:recapMac2}. One can check that the involved Jacobi theta functions in $\Theta_m(\tau,\bar\tau)$ satisfies transformation identities under the modular $T$ and $S$ transformations. These may introduce powers of $|\tau|$. These factors are compensated for by the corresponding transformations of $|\eta(\tau)|^2$. 

\paragraph{$B$-form codes}

There are a number of code equivalences which relate different codes to the same theory. These equivalences act as T-dualities at the level of the CFT. The result is that all real self-dual codes are equivalent to a code with the form
\begin{align}
    G^T \ = \ \left( \begin{array}{@{}c|c@{}}
   \begin{matrix}
      \alpha_1 \\
      \alpha_2 \\
      \vdots \\
      \alpha_n
   \end{matrix} 
      &
    \begin{matrix}
      \beta_1 \\
      \beta_2 \\
      \vdots \\
      \beta_n
   \end{matrix} 
\end{array} \right) \ = \ \left( \begin{array}{@{}c|c@{}}
   \begin{matrix}
      B
   \end{matrix} 
      &
    \begin{matrix}
      I
   \end{matrix} 
\end{array} \right)\, ,
\end{align}
where $B$ is an antisymmetric binary matrix. The generators~\eqref{eq:grecap} of codes in this form are simply given by
\begin{align}
    g_i = Z_i \prod_{j = 1}^n (X
    _j)^{B_{ij}}
\end{align}
Therefore every code theory can be specified by (at least) one binary antisymmetric $n\times n$ matrix $B$. Such matrices can be used to define undirected graphs with $n$ nodes and no self-links

\paragraph{Example: $n = 3$ code} Let us illustrate these elements with a simple example. Take
\begin{align}
    B = \begin{pmatrix}
    0 & 1 & 1 \\
    1 & 0 & 1 \\
    1 & 1 & 0
    \end{pmatrix} \, .
    \label{eq:Bexample}
\end{align}
\begin{figure}[h]
\centering
\includegraphics[width=0.2\textwidth]{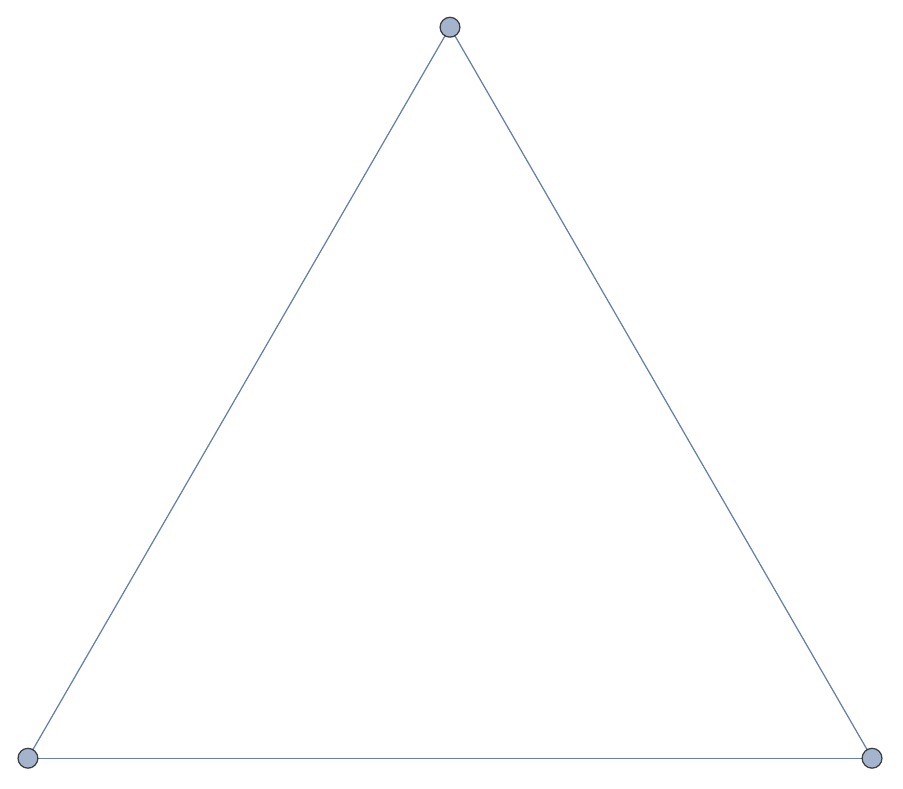}
\caption{Graph defining the matrix in equation~\eqref{eq:Bexample}.}
\label{fig:K3}
\end{figure}
This corresponds to the graph complete graph on 3 vertices (see figure \ref{fig:K3}).

We can use this to compute the generators:
\begin{align}
    g_1 \ &= \ Z_1 X_2 X_3\, , 
    \\
    g_2 \ &= \ X_1 Z_2 X_3 \, ,
    \\
    g_3 \ &= \ X_1 X_2 Z_3\, .
\end{align}
The full set of elements is then $G = \{ I,\, g_1,\, g_2,\, g_3,\, g_1 g_2,\, g_1 g_3,\, g_2 g_3,\, g_1 g_2 g_3 \}$. This allows us to compute the refined enumerator polynomial using~\eqref{eq:recapdefREP}. The result is
\begin{align}
    \tilde W_3(x_0, x_1, x_2) = x_0^3 + 3 x_0 x_1^2 + 4 x_2^3 \, .
    \label{eq:W3tilde}
\end{align}
This agrees with equation (6.21) of \cite{Dymarsky:2020qom}, and is equal to the $\tilde W_3$ used in the introduction. Here the tilde denotes that this is the extremal code at $n = 3$. We will return to this topic in section~\ref{sec:results}.

\section{Narain lattices and code theories at higher genus}
\label{sec:Higher-genus}

The goal of this section is to spell out the relation between quantum error-correcting codes and CFT partition functions at higher genus. Recall that a quantum error-correcting code can be specified by a collection of generators $g$ which satisfies certain properties. Each codeword is essentially a string of Pauli matrices, described by the symbols $\1,X,Y,Z$, which can also be specified by a pair of binary vectors $(\alpha, \beta)$. By relating the Pauli matrices to the elements of $\F_4=\{0,\omega,1,\omega^2\}$ via the Gray map, there is a relation to classical codes over $\F_4$.

In this section we will give a construction at higher-genus that is the exact counterpart of \eqref{eq:PFg1}. This will give a genus $g$ partition function of the form
\begin{equation}
    \label{eq:genusgformulaZ}
        Z^{(g)}(\Omega,\bar\Omega)=\frac{f_n\left( \Theta_{\vek m}(\Omega,\bar \Omega)\right)}{\Phi_g} \, .
    \end{equation}
Here $\Omega$ is the period matrix of the genus-$g$ Riemann surface, which is the direct higher-genus analog of the complex structure parameter $\tau$. $\Phi_g$  generalizes the contribution of  $2^n|\eta(\tau)|^{2n}$ in the genus 1 case and takes into account the contribution to the partition function due to oscillator modes. Formally, $\Phi_g= \text{det}^\prime \bar{\partial} $, which corresponds to the determinant of the Laplacian operator $ \bar{\partial} $ on the genus-$g$ Riemann surface with zero modes removed \cite{Knizhnik:1987xp}. This does not depend on the choice of Narain Lattice so it will not factor into our discussion.

Finally, $f_n$ is a degree $n$ homogeneous polynomial in $2^{g-1}(2^g+1)$ theta functions $\Theta_{\vek m}$. In this paper, we will relate $f_n(\Theta_{\vek m})$ to the \emph{refined higher-weight enumerator polynomial}, which we will define below.

\subsection{Higher-weight theta functions}

Given a Euclidean lattice $\Lambda$, it is natural to define the higher-weight lattice theta series 
\begin{equation}
    \Theta_{\Lambda}(\Omega) = \sum_{v_1\in\Lambda}\cdots\sum_{v_g\in \Lambda}\exp\left(2\pi i v_i\Omega_{ij}v_j\right)\, ,
\end{equation}
which is an analytic function of the Siegel upper half plane $\mathcal H_g$, defined by
\begin{equation}
    \mathcal H_g=\left\{\Omega\in\mathrm{Mat}_{g\times g}(\C)|\Omega=\Omega^T,\, \IM\Omega\succ 0\right\}\, ,
\end{equation}
with known modular transformation properties. In general, $\Theta_\Lambda$ will evaluate to a combination of higher-weight theta functions, see \cite{Igusa1972,Mumford1983,Mumford1983b}, \cite{AlvarezGaume:1986es}, or \cite{Oura2008}. Their general definition is through the sum
\begin{equation}
\label{eq:thetafunctiongen}
    \theta\left[\begin{matrix}\vek m\\\vek n\end{matrix}\right](\vek z,\Omega) =  \sum_{\vek k\in \Z^g} (-1)^{2(\vek k+\vek m)\cdot (\vek z+\vek n)}\exp\left(i\pi (\vek k+\vek m)\cdot\Omega(\vek k+\vek m)\right)\, ,
\end{equation}
where all quantities in bold font are length-$g$ (column) vectors. The theta functions \eqref{eq:thetafunctiongen} have well-known transformation properties under the genus $g$ modular maps, which will be discussed in detail in section~\ref{sec:relations}.

A number of the theta functions will be zero, generalizing the statement that $\theta_1(q) = 0$ in the genus 1 case. To see this at higher genus, we first introduce the notation
    \begin{equation}
    \label{eq:Thetamdef}
        \Theta_{\vek m}(\Omega,\bar \Omega)=\theta_{\vek m}(\Omega)\theta_{\vek m}(\Omega), \quad \vek m\in\mathsf{even}_g\, .
    \end{equation}
For $\Theta_{\vek m}$ in \eqref{eq:Thetamdef} to be non-zero, $\vek m$ ranges over the set of genus-$g$ even characteristics -- the subset of $\{1,2,3,4\}^g$ that contains an even number of ``1''s.
    There are $2^{g-1}(2^g+1)$ such even characteristics. For example, at genus 1 and 2, the theta functions are indexed by
    \begin{equation}
\mathsf{even}_1 = \{2,3,4\}\,,\qquad        \mathsf{even}_2=\{11,22,23,24,32,33,34,42,43,44\}\,.
    \end{equation}
The holomorphic function $\theta_{\vec m}(\Omega)$ in \eqref{eq:Thetamdef} is defined by
    \begin{equation}
\label{eq:genthetaconstants}
    \theta_{\vek m}(\Omega)=\theta\begin{bmatrix}\vek a(\vek m)\\\vek b(\vek m)\end{bmatrix}(0,\Omega)\, ,
\end{equation}
where $\vek a(\vek m)$ and $\vek b(\vek m)$ have entries according to
\begin{align}
    m_i=1&:\quad (a_i,b_i)=(\tfrac12,\tfrac12)\, ,&
    m_i=2&:\quad (a_i,b_i)=(\tfrac12,0)\, ,\\
    m_i=3&:\quad (a_i,b_i)=(0,0)\, ,
    &
    m_i=4&:\quad (a_i,b_i)=(0,\tfrac12)\, .
    \label{eq:miassignment}
\end{align}
For instance, at genus 1, we have
\begin{equation}
    \theta_2(\tau)=\theta\begin{bmatrix}
    1/2 \\ 0
    \end{bmatrix}(0,\tau)\, , \qquad
    \theta_3(\tau)=\theta\begin{bmatrix}
    0 \\ 0
    \end{bmatrix}(0,\tau)\, , \qquad
    \theta_4(\tau)=\theta\begin{bmatrix}
    0 \\ 1/2
    \end{bmatrix}(0,\tau)\, ,
\end{equation}
which are the usual Jacobi theta functions. 

\paragraph{Factorization limit}
In the analysis of higher-genus partition functions we will make use of the ``factorization limit,'' where a genus $g$ Riemann surface degenerates into two parts of genus $g-h$ and $h$ respectively, connected by an infinitely long thin tube. In this limit, $ \Omega$ becomes block-diagonal,
\begin{equation}
    \Omega_g \to \Omega_{g-h}\oplus \Omega_h\, ,
\end{equation}
and it is easy to verify that 
\begin{equation}
    \Theta_{a_1a_2\cdots a_g}(\Omega_g)\to \Theta_{a_1a_2\cdots a_g}( \Omega_{g-h}\oplus \Omega_h)=\Theta_{a_1\cdots a_{g-h}}(\Omega_{g-h})\Theta_{a_{g-h+1}\cdots a_g}(\Omega_{h})\, .
    \label{eq:factorsizationTheta}
\end{equation}

\subsection{From code to lattice theta series}
\label{subsec:derivation}

The lattice theta series is a function of the period matrices. At genus $g=1$, there is a definition of a theta function on a Lorentzian lattice $\Lambda$, defined by
\begin{equation}
    \Theta^{(1)}_\Lambda(q,\bar q)=\sum_{(\vec p_L,\vec p_R)\in \Lambda} q^{\frac{\vec p_L\cdot \vec p_L}2}\bar q^{\frac{\vec p_R\cdot \vec p_R}2}\, .
    \label{eq:thetagenus1}
\end{equation}
With this definition, the generalization of to genus 2 is 
\begin{equation}
    \Theta^{(2)}_\Lambda( q, \bar q, r, \bar r, s, \bar s)=\sum_{(\vec p_L,\vec p_R),(\vec k_L,\vec k_R)\in \Lambda}q^{\frac{\vec p_L\cdot \vec p_L}2}\bar q^{\frac{\vec p_R\cdot \vec p_R}2}r^{\vec p_L\cdot \vec k_L}\bar r^{\vec p_R\cdot \vec k_R}s^{\frac{\vec k_L\cdot \vec k_L}2}\bar s^{\frac{\vec k_R\cdot \vec k_R}2}\, .
\end{equation}
Here $q$, $r$, and $s$ are defined by
\begin{align}
    q = e^{2 i \pi \Omega_{11}},\ r = e^{2 i \pi \Omega_{12}},\ s = e^{2 i \pi \Omega_{22}}, \quad
   \bar q = e^{-2 i \pi\bar \Omega_{11}},\ \bar r = e^{-2 i \pi\bar \Omega_{12}},\ \bar s = e^{-2 i \pi\bar \Omega_{22}}\,.
\end{align}

\subsubsection{Derivation at genus 1}

Consider the (complete) genus 1 enumerator polynomial, introduced in \eqref{eq:genuPcomplete},
\begin{equation}
    W^{(1)}_{\mathcal C}(x_0,x_1,x_2,x_3)=\sum_{ c\in\mathcal C}x_0^{n-w_x(c)-w_y(c)-w_z(c)}x_1^{w_x( c)}x_2^{w_y( c)}x_3^{w_z( c)}\, .
\end{equation}
We will now write it in a different form, which is suitable for generalizations both to arbitrary fields and to higher genus,
\begin{equation}
    W^{(1)}_{\mathcal C}( x_{[\1]}, x_{[X]}, x_{[Y]}, x_{[Z]})=\sum_{\vec c\in \mathcal C}\prod_{i=1}^{n}  x_{[c_i]}\, .
\end{equation}
where $ x_{[\1]}=x_0$, $ x_{[X]}=x_1$, $ x_{[Y]}=x_2$ and $ x_{[Z]}=x_3$.
Each code-word $ c$ has entries $c_i$, taking the value in a size four set. We have given three equivalent formulations, where this set is
\begin{equation}
    \{\1,X,Y,Z\}\simeq \{0,\omega,1,\omega^2\}\simeq\{(0,0),(1,0),(1,1),(0,1)\}\, .
\end{equation}
The first of these equivalences correspond to relating the quantum code to a classical code over $\mathbb F_4$. The second corresponds to using the Gray map, and is what will be used when relating the quantum error-correcting code to the theta function of a Lorentzian lattice. We write
\begin{equation}
\label{eq:cvsalphabeta}
    c_i\  \leftrightarrow\ (\alpha_i,\beta_i)\, .
\end{equation}

The genus 1 theta series of the Lorentzian lattice $\Lambda(\mathcal C)$ associated to the code $\mathcal C$ by \eqref{eq:Lorentzianlatticefromcode}, is given by 
\begin{equation}
\label{eq:thetawithabsum}
    \Theta_{\Lambda(\mathcal C)}=\sum_{\vec c\in \mathcal C}\prod_{i=1}^n\sum_{a_i\in \Z}\sum_{b_i\in\Z}
    q^{\frac12(a_i+\frac{\alpha_i}2+b_i+\frac{\beta_i}2)^2}
    \bar q^{\frac12(a_i+\frac{\alpha_i}2-b_i-\frac{\beta_i}2)^2}\, ,
\end{equation}
where we substituted \eqref{eq:pLpR} into \eqref{eq:thetagenus1}.

Let us study the contribution in \eqref{eq:thetawithabsum} from a given $\vec c$ in the outermost sum and a given $i$ in the product. This corresponds to determining
\begin{equation}
\label{eq:Thetaofci}
    \Th( x_{[c_i]}):=\sum_{a_i\in \Z}\sum_{b_i\in\Z}
    q^{\frac12\left(a_i+\frac{\alpha_i}2+b_i+\frac{\beta_i}2\right)^2}
    \bar q^{\frac12\left(a_i+\frac{\alpha_i}2-b_i-\frac{\beta_i}2\right)^2}\, .
\end{equation}

We would like to simplify \eqref{eq:Thetaofci} so that it splits into two terms, such that in each term the two infinite sums separate. This is achieved by introducing the summation variables
\begin{equation}
    \mu=a_i+b_i,\quad \nu=a_i-b_i,
\end{equation}
and summing over even values of $ \mu+\nu$.   
The restriction to even $\mu+\nu$ can be implemented by inserting the ``complicated unit'' $\frac{1+(-1)^{\mu+\nu}}2$. With this substitution, 
\begin{align}
    \Th(x_{[c_i]})&=\frac12
    \bigg(\sum_{\mu\in\Z}q^{\frac12\left(\mu+\frac{\alpha_i+\beta_i}2\right)^2}\bigg)
    \bigg(\sum_{\nu\in\Z}\bar q^{\frac12\left(\nu+\frac{\alpha_i-\beta_i}2\right)^2}\bigg)
   \nonumber\\&\quad +
    \frac12
    \bigg(\sum_{\mu\in\Z}(-1)^\mu q^{\frac12\left(\mu+\frac{\alpha_i+\beta_i}2\right)^2}\bigg)
    \bigg(\sum_{\nu\in\Z}(-1)^\nu\bar q^{\frac12\left(\nu+\frac{\alpha_i-\beta_i}2\right)^2}\bigg)
    \label{eq:Thofciafterfactor}\, .
\end{align}
Depending on the values of $\alpha_i$ and $\beta_i$, each of the factor in \eqref{eq:Thofciafterfactor} evaluates to one of the Jacobi theta functions
\begin{align}
    (\alpha_i,\beta_i)&=(0,0) \, ,& \Th(x_{[\1]})&=\frac12\left(\theta_3(q)\theta_3(\bar q)+\theta_4(q)\theta_4(\bar q)\right)\, ,
    \\
    (\alpha_i,\beta_i)&=(1,0) \, ,& \Th(x_{[X]})&=\frac12\left(\theta_2(q)\theta_2(\bar q)+\theta_1(q)\theta_1(\bar q)\right)\, ,
    \\
    (\alpha_i,\beta_i)&=(1,1) \, ,& \Th(x_{[Y]})&=\frac12\left(\theta_3(q)\theta_3(\bar q)-\theta_4(q)\theta_4(\bar q)\right)\, ,
    \\
    (\alpha_i,\beta_i)&=(0,1) \, ,& \Th(x_{[Z]})&=\frac12\left(\theta_2(q)\theta_2(\bar q)-\theta_1(q)\theta_1(\bar q)\right)\, .
\end{align}
Now we can see why we introduced the refined enumerator polynomial with $ x_{[X]}= x_{[Z]}$, since $\Th( x_{[X]})=\Th( x_{[Z]})$ by the vanishing of $\theta_1(q)$.

\subsubsection{Derivation at higher genus}

It is natural to consider the higher-genus enumerator polynomial of a code over $F$,
\begin{equation}
\label{eq:higherweightenumeratorpolygen}
    W^{(g)}_{\mathcal C}(x_{[A]}):=\sum_{\mathbf M\in \mathcal C^g}\prod_{i=1}^n x_{[\mathrm{row}_i(\mathbf M)]}\, .
\end{equation}
This is a polynomial in $|F|^g$ variables $x_{[A]}$, where $A\in F^g$. 
Here the sum is over all possible $g$-tuples of codewords, packaged into the $n\times g$ matrix $\mathbf M$.

Having worked out the derivation of the genus 1 lattice theta series from a code, the generalization to higher-genus is straightforward.

The higher-weight enumerator polynomial \eqref{eq:higherweightenumeratorpolygen} is a sum over $g$-tuples of codewords, where each summand is a product over the entries indexed by $i$. We will now study each factor in such a product, i.e. for a fixed set of codewords $\vec c_{(1)},\ldots \vec c_{(g)}$ and index $i$. Define $\vek c_i$ with components $(c_{(1),i},\ldots,c_{(g)i})$, and further $\vek c_i\sim (\vek \alpha_i,\vek \beta_i)$ by \eqref{eq:cvsalphabeta}. Then
\begin{align}
    \Th(x_{[\vek c_i]})&=\sum_{\vek a\in \Z^g}\sum_{\vek b\in \Z^g}\exp\left(\frac{2\pi i}2\left(\vek a+\tfrac{\vek \alpha_i}2+\vek b+\tfrac{\vek \beta_i}2\right)\Omega\left(\vek a+\tfrac{\vek \alpha_i}2+\vek b+\tfrac{\vek \beta_i}2\right)\right)
    \nonumber\\
    &\qquad \qquad\quad \times \exp\left(\frac{2\pi i}2\left(\vek a+\tfrac{\vek \alpha_i}2-\vek b-\tfrac{\vek \beta_i}2\right)\bar\Omega\left(\vek a+\tfrac{\vek \alpha_i}2-\vek b-\tfrac{\vek \beta_i}2\right)\right)
    \, .
\end{align}
Now let $\vek \mu=\vek a + \vek b$ and $\vek \nu=\vek a-\vek b$, and write the sum as
\begin{equation}
    \sum_{\vek a\in \Z^g}\sum_{\vek b\in \Z^g}(\cdots)=\frac1{2^g}\sum_{\vek \mu\in\Z^g}\sum_{\vek \nu\in\Z^g}\prod_{h=1}^g\left(1+(-1)^{\mu_{(h)}+\nu_{(h)}}\right)(\cdots)
\end{equation}
It is now clear that the resulting expression,
\begin{align}
    \Th(x_{[\vek c_i]})&=\frac1{2^g}
    \sum_{\vek \mu\in\Z^g}\sum_{\vek \nu\in\Z^g}\exp\left(\frac{2\pi i}2\left(\vek \mu+\tfrac{\vek \alpha_i+\vek \beta_i}2\right)\Omega\left(\vek \mu+\tfrac{\vek \alpha_i+\vek \beta_i}2\right)\right)
    \nonumber\\
    &\qquad \qquad\quad \times  \exp\left(\frac{2\pi i}2\left(\vek \nu+\tfrac{\vek \alpha_i-\vek \beta_i}2\right)\bar\Omega\left(\vek \nu+\tfrac{\vek \alpha_i-\vek \beta_i}2\right)\right)\prod_{h=1}^g\left(1+(-1)^{\mu_{(h)}+\nu_{(h)}}\right)\, ,
\end{align}
will be a linear combination of the $\Theta_{\vek m}(\Omega,\bar \Omega)$ defined in \eqref{eq:Thetamdef} above.

In general, the variable $ x_{[\vek c_i]}$ for $\vek c_i\leftrightarrow (\vek\alpha_i,\vek \beta_i)$ maps to
\begin{equation}
    \Th(x_{[\vek c_i]})=\sum_{r_1}\sum_{r_2}\cdots \sum_{r_g}(-1)^{\sigma_1(r_1)+\sigma_2(r_2)+\ldots+\sigma_2(r_g)}\Theta_{r_1r_2\cdots r_g}(\Omega,\bar\Omega)\, ,
\end{equation}
where the sum over $r_h$ is determined by the following
\begin{align}
    (\alpha_{(h),i},\beta_{(h),i})&=(0,0) & r_h&=3,4,\quad \sigma_h(3)=\sigma_h(4)=0\, ,
    \\
    (\alpha_{(h),i},\beta_{(h),i})&=(1,0) & r_h&=2,1,\quad \sigma_h(2)=\sigma_h(1)=0\, ,
    \\
    (\alpha_{(h),i},\beta_{(h),i})&=(1,1) & r_h&=3,4,\quad \sigma_h(3)=0,\ \sigma_h(4)=1\, ,
    \\
    (\alpha_{(h),i},\beta_{(h),i})&=(0,1) & r_h&=2,1,\quad \sigma_h(2)=0,\ \sigma_h(1)=1\, .
\end{align}

\subsubsection{Explicit formulas at genus 2}

Like the case at genus 1, some of the variables in the complete enumerator polynomial map to identical theta functions. At genus two, there are ten non-zero theta functions $\Theta_{\vek m}$, while six theta functions identically vanish: $\Theta_{12}=\Theta_{13}=\Theta_{14}=\Theta_{21}=\Theta_{31}=\Theta_{41}=0$. Taking this into account, we find that
\begin{align}
 y_0  :=   x_{[\1\1]}  &\ \mapsto \  \Theta_{33}+\Theta_{34}+\Theta_{43}+\Theta_{44}  
 \, ,   \\
 y_1  :=  x_{[\1 Y]}  &\ \mapsto \  \Theta_{33}-\Theta_{34}+\Theta_{43}-\Theta_{44}  
  \, ,  \\
   y_2:=  x_{[Y \1] }  &\ \mapsto \  \Theta_{33}+\Theta_{34}-\Theta_{43}-\Theta_{44} 
\, ,   \\
 y_3:= x_{[YY]}  &\ \mapsto \  \Theta_{33}-\Theta_{34}-\Theta_{43}+\Theta_{44} 
 \, ,  \\
    y_4:=  x_{[\1 X]}  =  x_{[\1 Z]}  &\ \mapsto \  \Theta_{32}+\Theta_{42} 
  \, ,  \\
   y_5:=  x_{[X \1]}  = 
     x_{[Z \1]}  &\ \mapsto \ 
     \Theta_{23}+\Theta_{24}
 \, ,   \\
y_6:= x_{[X X]}  =   x_{[ZZ]} &\ \mapsto \ 
     \Theta_{11}+\Theta_{22}
 \, ,    \\
 y_7:= x_{[Y X ]} =    x_{[Y Z]}  &\ \mapsto \  \Theta_{32}-\Theta_{42}
 \, ,    \\
    y_8:= x_{[X Y]}  =   x_{[Z Y]}  &\ \mapsto \  \Theta_{23}-\Theta_{24} 
 \, ,    \\
  y_9:= x_{[X Z]} =  x_{[Z Z]} &\ \mapsto \  -\Theta_{11}+\Theta_{22} 
  \, .
\end{align}
We see that compared to the original 16 code variables $x_{[c_ic'_i]}$, we are now considering a subspace spanned by ten variables $y_i$. This is the genus 2 version of going from the complete enumerator polynomial to the refined enumerator polynomial.

One may also write an expression that gives the genus two enumerator polynomial directly in terms of the $y_i$, but the resulting formula is not particularly illuminating:
\begin{align}
    \nonumber
W_{g=2}(y_0,\ldots ,y_9)&=\sum_{c\in\mathcal C}\sum_{\tilde c\in\mathcal C}
y_0^{|(1-\alpha)\land(1-\beta)\land(1-\tilde\alpha)\land(1-\tilde\beta)|}
y_1^{|(1-\alpha)\land(1-\beta)\land\tilde\alpha\land\tilde\beta|}
y_2^{|\alpha\land\beta\land(1-\tilde\alpha)\land(1-\tilde\beta)|}
\nonumber\\&\quad\quad\quad\quad\times
y_3^{|\alpha\land\beta\land\tilde\alpha\land\tilde\beta|}
 y_4^{|(1-\alpha)\land(1-\beta)\land \tilde\alpha\land(1-\tilde\beta)|+|(1-\alpha)\land(1-\beta)\land(1-\tilde\alpha)\land \tilde\beta|}
 \nonumber\\&\quad\quad\quad\times
y_5^{|\alpha\land(1-\beta)\land(1-\tilde\alpha)\land(1-\tilde\beta)|+|(1-\alpha)\land\beta\land(1-\tilde\alpha)\land(1-\tilde\beta)|}
\nonumber\\&\quad\quad\times 
y_6^{|\alpha\land(1-\beta)\land\tilde\alpha\land(1-\tilde\beta)|+|(1-\alpha)\land\beta\land(1-\tilde\alpha)\land\tilde\beta|}
y_7^{|\alpha\land\beta\land\tilde\alpha\land(1-\tilde\beta)|+|\alpha\land\beta\land(1-\tilde\alpha)\land\tilde\beta|}
\nonumber\\&\quad\times 
 y_8^{|\alpha\land(1-\beta)\land\tilde\alpha\land\tilde\beta|+|(1-\alpha)\land\beta\land\tilde\alpha\land\tilde\beta|}
y_9^{|(1-\alpha)\land\beta\land \tilde\alpha\land(1-\tilde\beta)|+|\alpha\land(1-\beta)\land(1-\tilde\alpha)\land\tilde\beta|}
\label{eq:Wgenus2def}\, ,
\end{align}
where $\land$ denotes the component-wise ``and'' operator.

We are also interested in the factorization limit \eqref{eq:factorsizationTheta} in terms of code variables. For the case of genus 2, it takes the form
\begin{align}
    y_0&\mapsto x_0x'_0
    \,,&
y_1&\mapsto x_0x'_1
    \,,&
y_2&\mapsto x_1x'_0
    \,,&
y_3&\mapsto x_1x'_1
    \,,&
y_4&\mapsto x_0x'_2 \nonumber
    \,,\\
y_5&\mapsto x_2x'_1
    \,,&
y_6&\mapsto x_2x'_2
    \,,&
y_7&\mapsto x_1x'_2
    \,,&
y_8&\mapsto x_2x'_1
    \,,&
y_9&\mapsto x_2x'_2
    \,,
    \label{eq:factorizationytox}
\end{align}
where the $x_i$ refer to variables on the left genus 1 Riemann surface and $x'_i$ to the variables on the right genus 1 Riemann surface.

%
%

\subsection{From theta relations to polynomial relations}
\label{sec:relations}

The goal now will be to study the transformation properties of the higher-weight theta functions in order to determine how they lift to transformations of the enumerator polynomials. The modular transformations are
\begin{equation}\label{eq: modular tranformation general genus}
    \Omega\mapsto \Omega'=(A\Omega+B)(C\Omega+D)^{-1}, \qquad \Gamma=\begin{pmatrix}A&B\\C&D
    \end{pmatrix}\in \mathrm{Sp}(2g,\mathbb Z)\, . 
\end{equation}
The corresponding transformations of the higher-genus theta functions is given by \cite{Igusa1972} (see also e.g. \cite{AlvarezGaume:1986es})
\begin{equation}
\label{eq:thetaafter}
\theta\begin{bmatrix}\vek a'\\ \vek b'\end{bmatrix}(0,\Omega')=\epsilon(\Gamma)\exp(-i \pi\phi(\vek a,\vek b,\Gamma))\sqrt{\det(C\Omega+D)}\, \theta\begin{bmatrix}\vek a\\ \vek b\end{bmatrix}(0,\Omega)\, ,
\end{equation}
where
\begin{equation}
    \begin{pmatrix}
    \vek a'\\\vek b'
    \end{pmatrix}=\begin{pmatrix}
    D & -C \\ -B & A
    \end{pmatrix}\begin{pmatrix}\vek a\\\vek b
    \end{pmatrix}+\frac12\begin{pmatrix}
    (CD^T)_{\mathrm{diag}}\\(AB^T)_{\mathrm{diag}}\\
    \end{pmatrix}\, ,
\end{equation}
where $\epsilon(\Gamma)$ is a phase which is an eighth root of unity, $\phi(\vek a,\vek b,\Gamma)=\vek a.D^TB\vek a+\vek b.C^TA\vek b-2\vek a.B^TC\vek b+(\vek a.D^T-\vek b.C^T)(AB^T)_{\mathrm{diag}}$, and $M_{\mathrm{diag}}$ denotes the diagonal entries of a matrix $M$, seen as a column vector.

In our construction, the theta functions always come in pairs $\theta(\Omega) \theta(\bar \Omega)$, so any phases appearing from the modular transformations will cancel between the holomorphic and antiholomorphic parts. Furthermore, under the modular transformations \ref{eq: modular tranformation general genus}, the denominator $|\Phi_g|$ transforms covariantly with modular weights $(\frac{c}{2},\frac{c}{2})$. This is required to cancel the square-root factor in \eqref{eq:thetaafter}.

Consider now the set of $\mathrm{Sp}(2g,\mathbb Z)$ transformation acting on the theta functions $\varTheta_{\vek m}$. Up to the weights imposed by the square-root factor in \eqref{eq:thetaafter}, such transformations amount to mapping the $\varTheta_{\vek m}$ among each other. We now wish to lift these relations to the polynomial variables. It is clear that we only need to exhibit this lift for the generators of $\mathrm{Sp}(2g,\mathbb Z)$.

As a warm-up, consider the case $g=1$. We can take as generators
\begin{equation}
    T: \quad \tau\mapsto \tau+1,\qquad S: \tau\mapsto -\frac1\tau\,.
\end{equation}
The corresponding transformations of the Jacobi theta functions induce 
\begin{equation}
    T:\quad \begin{cases}
    \Theta_2(\tau+1)=\Theta_2(\tau)\,,
    \\
    \Theta_3(\tau+1)=\Theta_4(\tau)\,,
    \\
    \Theta_4(\tau+1)=\Theta_3(\tau)\,,
    \end{cases}
 \qquad S:\quad \begin{cases}
    \Theta_2(-1/\tau)=|\tau|\Theta_4(\tau)\,,
    \\
    \Theta_3(-1/\tau)=|\tau|\Theta_3(\tau)\,,
    \\
    \Theta_4(-1/\tau)=|\tau|\Theta_2(\tau)\,.
    \end{cases}
\end{equation}
These relations lift to
\begin{equation}
    T:\quad \begin{cases}
    x_0\mapsto x_0\,,
    \\
x_1\mapsto- x_1\,,
    \\
    x_2\mapsto x_2\,,
    \end{cases}
     \qquad S:\quad \begin{cases}
     x_0 \mapsto \frac{1}{2}(x_0 + x_1 + 2x_2) \,,
     \\
     x_1 \mapsto  \frac{1}{2}(x_0 + x_1 - 2x_2)\,,
     \\
      x_2 \mapsto \frac{1}{2} (x_0-x_1)\,,
    \end{cases}
\end{equation}
which are exactly the MacWilliams identities \eqref{eq:MacWilliamsG1} and \eqref{eq:MacWilliamsExtra} given in section~\ref{sec:review}.

To generalize to higher genus, it is convenient to use the generators of $\mathrm{Sp}(2g,\Z)$ as given by \cite{Stanek1963} and reviewed in \cite{Henriksson:2021qkt}. At genus $g=2$ and $g=3$, there are three generators; in all other cases there are only two. At genus 2, the generators can be taken to be $T$, $R$, and $D$, with the corresponding matrices $\Gamma$ of the form
\begin{align}
        T :\ \Gamma= \begin{pmatrix}
        1 & 0 & 1 & 0\\
        0 & 1 & 0 & 0\\
        0 & 0 & 1 & 0\\
        0 & 0 & 0 & 1
        \end{pmatrix}  \, ,\quad
        R :\quad \Gamma
        = \begin{pmatrix}
        1 & 0 & 0 & 0\\
        1 & 1 & 0 & 0\\
        0 & 0 & 1 & -1\\
        0 & 0 & 0 & 1
        \end{pmatrix}  \, ,  \quad 
        D :\ \Gamma= \begin{pmatrix}
        0 & 1 & 0 & 0\\
        0 & 0 & -1 & 0\\
        0 & 0 & 0 & 1\\
        1 & 0 & 0 & 0
        \end{pmatrix}\, .
\end{align}
These are related to the more familiar set of genus 2 generators, $T_1$, $T_2$, $U$, $S_1$, $S_2$, by the following: 
\begin{equation}
    T_1 = T\, , \quad T_2=D^{-1}TD,\quad U=DRD^{-1}\, ,\quad S_1=TD^2TD^2T\, , \quad S_2=DRD^2RD^2R\, ,
\end{equation}
where $D^{-1}=D^7$.
Using \eqref{eq:thetaafter}, the transformations of the theta functions $\Theta_m(\Omega,\bar\Omega)$ can be determined, and the corresponding equations for the polynomial variables are

\begin{align}
\label{eq:modulargenus2}
    T_1:\ \begin{cases}
    y_2\mapsto -y_2
  \, ,  \\
    y_3\mapsto -y_3
  \, ,  \\
    y_7\mapsto -y_7\, ,
    \end{cases}
     \quad     T_2:\ \begin{cases}
    y_1\mapsto -y_1
 \, ,   \\
    y_3\mapsto -y_3
  \, ,  \\
    y_8\mapsto -y_8\, ,
    \end{cases}
     \quad     U:\ \begin{cases}
    y_7\mapsto -y_7
  \, ,  \\
    y_8\mapsto -y_8
\, ,    \\
    y_9\mapsto -y_9\, ,
    \end{cases} \\
    \label{eq:modulargenus2-more}
    S_1:\ \begin{cases}
    y_0\mapsto \frac{1}{2}\left(y_0 + y_2 + 2 y_5\right)
\, ,    \\
    y_1\mapsto \frac{1}{2}\left(y_1 + y_3 + 2 y_8\right)
 \, ,   \\
    y_2\mapsto \frac{1}{2}\left(y_0 + y_2 - 2 y_5\right)
 \, ,   \\
    y_3\mapsto  \frac{1}{2}\left(y_1 + y_3 - 2 y_8\right)
  \, ,  \\
    y_4\mapsto \frac{1}{2}\left(y_4 + y_6 + y_7 + y_9\right)
 \, ,   \\
    y_5\mapsto \frac{1}{2}\left(y_0 - y_2\right)
 \, ,   \\
    y_6\mapsto \frac{1}{2}\left(y_4 + y_6 - y_7 - y_9\right)
  \, ,  \\
    y_7\mapsto \frac{1}{2}\left(y_4 - y_6 + y_7 - y_9\right)
  \, ,  \\
    y_8\mapsto \frac{1}{2}\left(y_1 - y_3\right)
 \, ,   \\
    y_9\mapsto \frac{1}{2}\left(y_4 - y_6 - y_7 + y_9\right)
    \end{cases}
     \quad     S_2:\ \begin{cases}
    y_0\mapsto \frac{1}{2}\left(y_0 + y_1 + 2 y_4\right)
  \, ,  \\
    y_1\mapsto \frac{1}{2}\left(y_0 + y_1 - 2 y_4\right)
  \, ,  \\
    y_2\mapsto \frac{1}{2}\left(y_2 + y_3 + 2 y_7\right)
 \, ,   \\
    y_3\mapsto \frac{1}{2}\left(y_2 + y_3 - 2 y_7\right)
  \, ,  \\
    y_4\mapsto  \frac{1}{2}\left(y_0 - y_1\right)
 \, ,   \\
    y_5\mapsto \frac{1}{2}\left(y_5 + y_6 + y_8 + y_9\right)
  \, ,  \\
    y_6\mapsto \frac{1}{2}\left(y_5 + y_6 - y_8 - y_9\right)
 \, ,   \\
    y_7\mapsto  \frac{1}{2}\left(y_2 - y_3\right)
\, ,    \\
    y_8\mapsto \frac{1}{2}\left(y_5 - y_6 + y_8 - y_9\right)
 \, ,   \\
    y_9\mapsto \frac{1}{2}\left(y_5 - y_6 - y_8 + y_9\right)
    \, .
    \end{cases}
\end{align}


\section{Partition functions for code theories}
\label{sec:results}

In the previous section, we have described how code CFTs are constrained by modular invariance for general genus, and by factorization limits, which relate different genera. Now we shall explicitly demonstrate how to use these requirements to constrain the space of possible code theories. This is essentially a primitive example of the modular bootstrap -- because modular invariance is so simple for the code theories, we can enumerate all of its possible solutions. Genus 1 modular invariance completely determines all of the $n = 1$ and $n = 2$ code theories. We shall see that genus 2 considerations are enough to fix the space of $n = 3$ code theories, so this will be our primary example, given in section \ref{subsec:explicitN3}. At $n> 3$, genus 2 constraints greatly reduce the space of theories but do not entirely fix them. We will summarize the classification of invariant polynomials for general $n$ in section \ref{subsec:invar}, and for $n \leqslant6$ we will compare the number of valid (modular invariant polynomial with positive integer coefficients) genus 1 partition functions, genus 2 partition functions, and actual code theories in section \ref{subsec:counting}.

\subsection{Finding invariant polynomials: example at $n = 3$}
\label{subsec:explicitN3}

\begin{figure}[t]%
    \centering
    \subfloat[\centering]{{\includegraphics[width=7cm]{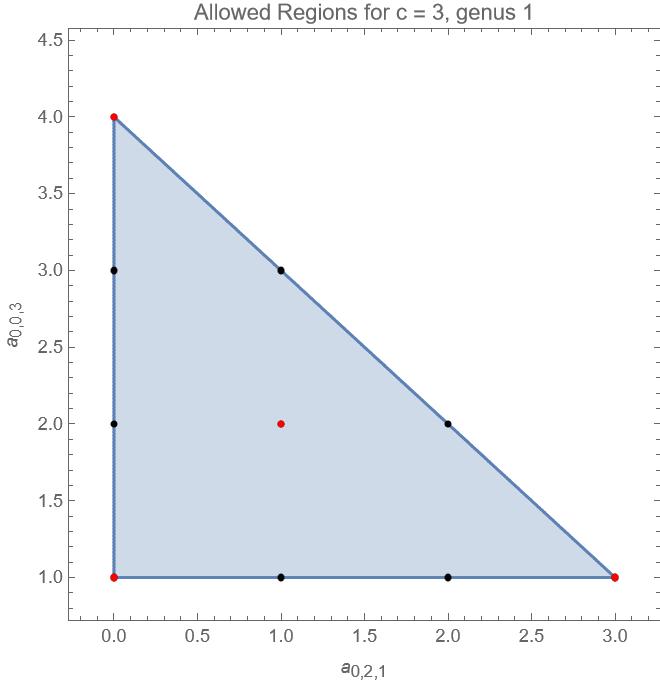} }}%
    \qquad
    \subfloat[\centering ]{{\includegraphics[width=7cm]{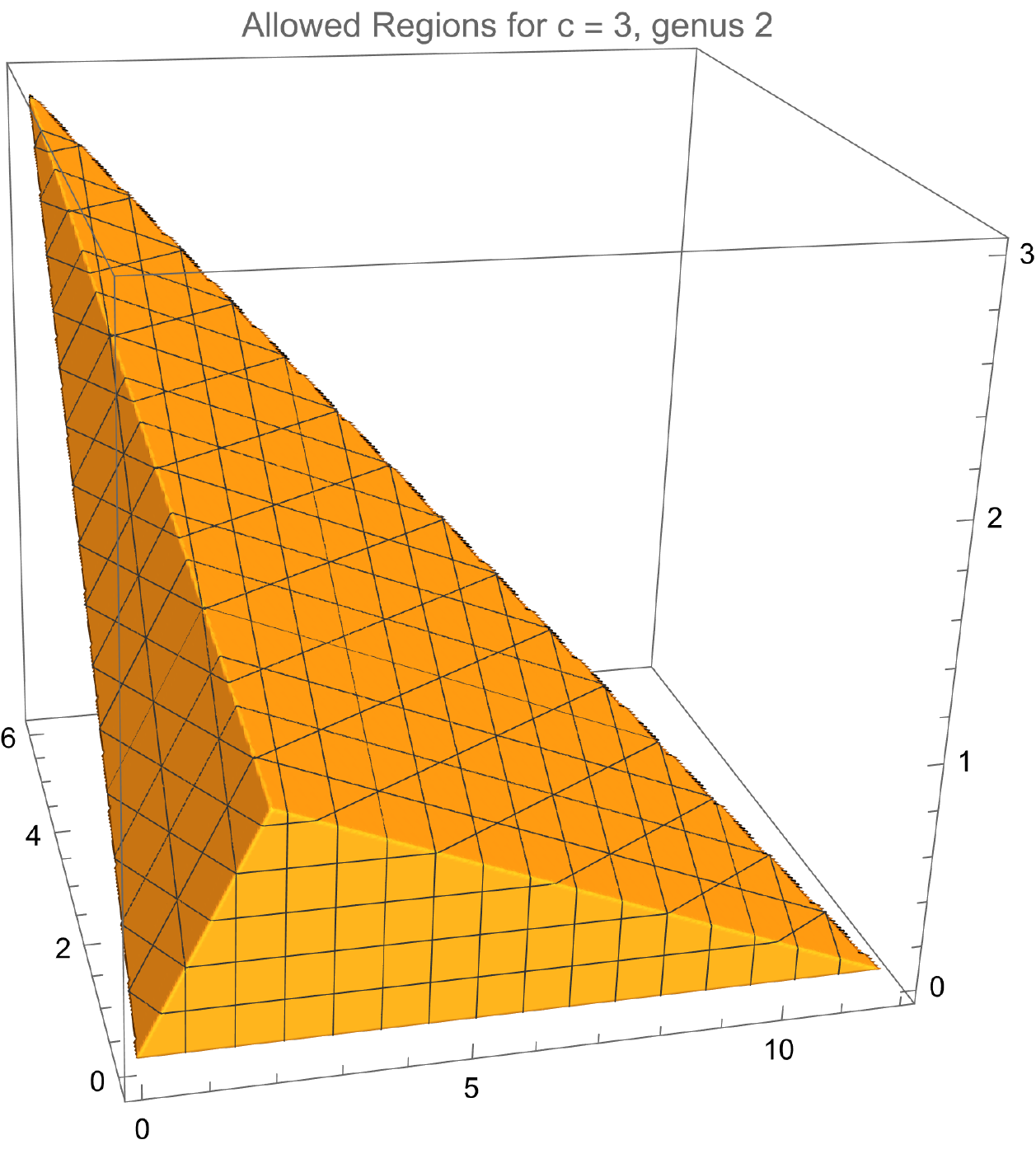} }}%
    \caption{Allowed regions for $n = 3$. Plot (a): 2d region for genus 1.  Dots represent polynomials with integer coefficients. Red dots, appearing at $(0,1)$, $(3, 1)$, $(1,2)$, and $(0,4)$, represent factorization limits of genus 2 polynomials, as well as physical code theories. Plot (b): rough 3d plot of genus 2 allowed region. Remarkably, the factorizing, positive integer solutions lie on the vertices of this polytope! 
    }%
    \label{fig:cequals3}%
\end{figure}

Let us start by presenting the case $n = 3$ in full detail. For genus 1, the most general homogeneous degree 3 polynomial is
\begin{align}
    P^{(1)}_{\mathrm{gen}}[3] = x_0^3 + a_{2,1,0} x_0^2 x_1 +  a_{2,0,1} x_0^2 x_2 + \ldots\, ,
\end{align}
which has 10 terms in total. Now recall that for genus 1, modular invariance implies that the polynomial is invariant under
\begin{align}
\begin{split}
    S:& \qquad x_0 \to \frac{1}{2}(x_0 + x_1 + 2x_2) \,, \quad x_1 \to \frac{1}{2}(x_0 + x_1 - 2x_2) \,, \quad x_2 \to \frac{1}{2}(x_0 - x_1) \,, \\
    T:& \qquad x_1 \to -x_1 \,.
\end{split}\label{eq: n3sec genus 1 transformations}
\end{align}
This fixes all but two of the undetermined coefficients, so we are left with
\begin{align}
\begin{split}
    & P^{(1)}_{\mathrm{inv}}[3] = x_0^3 + (4-a_{0,0,3}) x_0 x_2^2 + (-1 + a_{0,0,3}) x_0 x_1^2 \\
    & \qquad \qquad \qquad + (4 - a_{0,0,3} - a_{0,2,1}) x_0^2 x_2 + a_{0,2,1} x_1^2 x_2 + a_{0,0,3} x_2^3  \, .
\end{split}
\end{align}
Requiring that all of these coefficients are positive leads to the inequalities
\begin{align}
    0 \leq a_{0,2,1} \leq 3, \qquad 1 \leq a_{0,0,3} \leq 4 - a_{0,2,1} \, , 
\end{align}
which can be displayed as the two-dimensional region shown in figure~\ref{fig:cequals3}(a). Finally, we require that each of these coefficients is an integer, which leads to 10 solutions, the dots in figure~\ref{fig:cequals3}(a).

\vspace{5mm}
Now consider genus two. We again write the most general polynomial and then impose modular invariance. The result is a large polynomial
\begin{align}
\begin{split}
    & P^{(2)}_{\mathrm{inv}}[3] = y_0^3 + a_{0,0,0,0,0,0,0,1,1,1} y_3 y_7 y_8 + \ldots \, , 
\end{split}
\end{align}
which has a total of 48 terms and 3 undetermined parameters. Requiring positivity of all coefficients again gives a finite region, which we have displayed in figure~\ref{fig:cequals3} (b). If we require that the coefficients are all integers, we find 11 solutions. So there are 11 potential genus 2 partition functions.

However code theories must have genus 2 partition functions which factorize into genus 1 partition functions in the limit where the genus 2 Riemann surface becomes degenerate. Recall that this limit gives
\begin{align}
    y_0&\mapsto x_0x'_0
    \,,&
y_1&\mapsto x_0x'_1
    \,,&
y_2&\mapsto x_1x'_0
    \,,&
y_3&\mapsto x_1x'_1
    \,,&
y_4&\mapsto x_0x'_2 \nonumber
    \,,\\
y_5&\mapsto x_2x'_1
    \,,&
y_6&\mapsto x_2x'_2
    \,,&
y_7&\mapsto x_1x'_2
    \,,&
y_8&\mapsto x_2x'_1
    \,,&
y_9&\mapsto x_2x'_2
    \,.
\end{align}
Doing this, one finds that only four factorize of the genus 2 polynomials factorize into genus 1 polynomials, and in fact these are precisely the four the genus 1 polynomials which come from actual codes via New Construction A, given in \eqref{eq:intro code theories1}-\eqref{eq:intro code theories4}! The factorizing polynomials, which correspond to the set of real theories in this case, are displayed in red in figure~\ref{fig:cequals3} (a). They sit at the vertices of the polytope in figure~\ref{fig:cequals3} (b).

This is the general procedure we will follow for $n = 3, 4, 5, 6$ in the later sections. Actually, the case of $n = 3$ is a little special. We could have determined all real theories by only requiring factorization, without positivity at genus 1 or genus 2. We believe this is related to the fact that at $n = 3$, genus 2 is enough to eliminate all fake theories -- essentially, it fully fixes the partition function (given the assumption that it has enumerator polynomial form). We conclude that at $n = 3$, there are no non-code theories with EP form because only the four code theories are consistent with factorization at genus 2. 

At $n = 4$, where genus 2 does not remove all of the fake theories, factorization without positivity is not enough to find all of the modular invariant polynomials. Furthermore, just like $n = 3$ at genus 1, the real $n = 4$ theories do not all lie at the vertices of the genus 1 and genus 2 polytopes; some lie on the edges, faces, and in the bulk. Thus we find two special properties obeyed by $n = 3$ at genus 2: 1) the set of real theories is fixed only by factorization, and 2) the real theories lie on the corners of the polytope. As such, it would be very interesting to push our calculation to genus 3, to see if these two patterns persist, but we leave this for future work.

\subsection{Invariant polynomial ring}
\label{subsec:invar}

Now let us be more general, and describe the method of classifying invariant polynomials, valid for any $n$. We will be interested in the ring $R_g$ of polynomials invariant under the genus $g$ modular relations. $R_g$ admits a grading by degree
\begin{equation}
    R_g=\bigoplus_{n=0}^\infty R_g^{[n]}\, ,
\end{equation}
where $\dim R_g^{[n]}$ can be computed by Molien's formula, which we will now review (see e.g. \cite{Stanley1979}). For a matrix group $G$ acting on $n$ variables $x_0,\ldots x_{n-1}$, Molien's formula gives the dimension of the space $R_g^{[n]}$ of degree $n$ invariant polynomials

\begin{align}
\label{eq:Molienseriesformula}
    M(r)   = \sum_{n=0}^\infty r^n\dim(R_g^{[n]}) = \frac{1}{|G|} \sum_{g \in G} \det(1 - rg)^{-1} \, ,
\end{align}
where $|G|$ is the number of (distinct) elements in $G$, and $g$ are the individual matrices in the representation in question. 

\paragraph{Genus 1:} Recall that invariance under the modular group $SL(2, \Z) = SP(2, \mathbb{Z})$ acting on the three-dimensional vector space of $\{x_0, x_1, x_2\}$, is given by equation~\eqref{eq: n3sec genus 1 transformations}. Using \eqref{eq:Molienseriesformula}, we find that Molien series for this group acting on 3 variables is given by
\begin{align}
\label{eq:Moliengenus1}
    M(r) = \frac{1}{(1-r)(1-r^2)(1-r^3)} = 1 + r + 2r^2 + 3r^3 + 4r^4 + 5r^5 + 7r^6 + 8 r^7 + \ldots\, .
\end{align}
The number of solutions to the transformations \eqref{eq: n3sec genus 1 transformations} for general degree-$n$ homogeneous polynomials is given by the $r^n$ coefficient in the Molien series in~\eqref{eq:Moliengenus1}. This lets us infer the number of new generators at each order. The whole ring $R_1$ is generated by three polynomials $p^{(g=1)}_n(x_0,x_1,x_2)$, of degree $n$ for $n=1,2,3$ \cite{Dymarsky:2020bps}. 

\paragraph{Genus 2:} 
At genus 2, we have a 10-dimensional representation of $SP(4, \mathbb{Z})$ furnished by the 10 $y$-variables. The transformations are given in~\eqref{eq:modulargenus2}--\eqref{eq:modulargenus2-more} and we find that $|G| = 720$. This allows us to calculate the Molien series
\begin{align}
\label{eq:Moliengenus2}
   M(r) = \frac{f(r)}{(1-r)^2(1-r^2)(1-r^3)^2(1-r^4)^2(1-r^5)^2(1-r^6)}\, ,
\end{align}
where 
\begin{align}\begin{split}
    f(r)&=1 + r + r^4 + r^5 + 3 r^6 + 2 r^7 + 3 r^8 + 3 r^9 + 5 r^{10} + 3 r^{11} + 10 r^{12} + 6 r^{13} \\
    &\quad + 9 r^{14} + 7 r^{15} + 4 r^{16} + 2 r^{17} + r^{19}\, .
    \end{split}
\end{align}
The number of generators is calculated using the Molien series \ref{eq:Moliengenus2}. In principle, there could appear non-trivial relations between the generators. We find that there are no non-trivial relations among the terms constructed from generators at lower degree. Therefore, the generators organize as given in table~\ref{tab:genus2generators}.

\begin{table}
\begin{center}
\caption{Number of linearly independent invariant polynomials, and number of generators of degree $n$ at genus $g=2$.}\label{tab:genus2generators}
\begin{tabular}{ c | c c c c c c c c c c c c}
 $n$ & 1 & 2 & 3 & 4 & 5 & 6 & 7 & 8 & 9 & 10 & 11 & 12\\ 
 \hline
 invariant polynomials & 1 & 2 & 4 & 8 & 14 & 27 & 46 & 82 & 140 & 237 & 386 & 630\\  
 new generators & 1 & 1 & 2 & 3 & 4 & 6 & 7 & 9 & 11 & 10 & 4 & 0
\end{tabular}
\label{tab:genus2generators}
\end{center}
\end{table}

There are new generators only up to $n = 11$. A general method to completely specify this full set is to use enumerator polynomials of actual code theories as a set of generators. At each $n$ from $n=1$ to $n=11$ one can choose the new generators arising at each $n$ in Table~\ref{tab:genus2generators} to be those from actual (inequivalent) code theories.

\subsection{Finding invariant polynomials: results for various $n$}
\label{subsec:counting}

Now let us describe the results of counting invariant polynomial for general values of $n$. There are too many genus 2 solutions to count above $n = 6$, but in principle our method will work for any order $n$.  

There are 58 generators, as described in table~\ref{tab:genus2generators} above. Rather than list all of them, we will list the full set of adjacency graphs associated with matrices $B$ which can be used to calculate the complete ring of invariant polynomials using \eqref{eq:Wgenus2def}. The set of generators is given in appendix \ref{app:gengraphs}.

\paragraph{$\boldsymbol{n = 1}$}

At genus 1, we have a single polynomial:
\begin{align}
    W_1 = x_0 + x_2 \, .
\end{align}
At genus 2, we also find one polynomial,
\begin{align}
    W_1^{(2)} = y_0 + y_4 + y_5 + y_6 \, ,
\end{align}
which reproduces $W_1(x_0,x_1,x_1)W_1(x'_0,x'_1,x'_2)$ in the factorization limit, \eqref{eq:factorizationytox}. These polynomials correspond to the single unique code with $B=0$, or equivalently the graph with one node and zero vertices.

\paragraph{$\boldsymbol{n = 2}$}

At genus 1, we have two polynomials:
\begin{align}
\begin{split}
    W_2 \ &= \ x_0^2 + x_1^2 + 2 x_2^2 \\
    ( W_1 )^2 \ &= \ (x_0 + x_2)^2
\end{split}
\end{align}

There are two genus 2 polynomials which reproduce each of these polynomials in the factorization limit. They are
\begin{align}
\begin{split}
    W^{(2)}_2 \ &= \ y_0^2 + y_1^2 + y_2^2 + y_3^3 + 2(y_4^2 + y_5^2 + y_6^2 + y_7^2 + y_8^2 + y_9^2)
    \, ,\\
    (W^{(2)}_1)^2 \ &= \ (y_0 + y_4 + y_5 + y_6)^2 \, .
\end{split}
\end{align}
These polynomials correspond to two unique codes corresponding to 
\begin{equation}
B=\begin{pmatrix}
     0 & 1\\
     1 & 0
\end{pmatrix}
\end{equation}

and 

\begin{equation}
B=\begin{pmatrix}
     0 & 0\\
     0 & 0
\end{pmatrix}\, .
\end{equation}
These polynomials can be taken to be the new generators at this order. 
 
 \paragraph{$\boldsymbol{n = 3}$}

At genus 1, we have 10 polynomials, out of which only 4 arise from in-equivalent codes.

At genus 2, there are 11 polynomials but only 4 factorize. There are only 4 $B$-form codes, so the set of codes is entirely determined by consistency with genus 2 modular invariance. 

Since there are 2 new generators at this order, we can choose the following code generator matrices to define them: 
\begin{equation}
B=\begin{pmatrix}
    0 & 1 & 1\\
    1 & 0 & 1\\
    1 & 1 &0
\end{pmatrix} , \qquad 
B=\begin{pmatrix}
    0 & 1 & 0\\
    1 & 0 & 1\\
    0 & 1 & 0
\end{pmatrix}
\end{equation}
to get

\begin{align*}
W^{(2)}_{3}=&y_{0}^3 + 3 y_{1}^2 y_{4} + 3 y_{0} y_{4}^2 + y_{4}^3 + 3 y_{2}^2 y_{5} + 
 3 y_{0} y_{5}^2 + y_{5}^3 + 3 y_{3}^2 y_{6} + 6 y_{4} y_{5} y_{6}+ 3 y_{0} y_{6}^2 + y_{6}^3 + 3 y_{5} y_{7}^2  \\
& + 3 y_{6} y_{7}^2 + 6 y_{3} y_{7} y_{8} + 3 y_{4} y_{8}^2 + 3 y_{6} y_{8}^2+ 6 y_{2} y_{7} y_{9} + 6 y_{1} y_{8} y_{9} + 3 y_{4} y_{9}^2 + 3 y_{5} y_{9}^2 \,,  \\
\tilde{W}^{(2)}_{3}=&y_{0}^3 + 3 y_{0} y_{1}^2 + 3 y_{0} y_{2}^2 + 6 y_{1} y_{2} y_{3} + 
 3 y_{0} y_{3}^2 + 4 y_{4}^3+ 4 y_{5}^3 + 4 y_{6}^3 + 12 y_{4} y_{7}^2 + 12 y_{5} y_{8}^2 + 12 y_{6} y_{9}^2 \, .
\end{align*}
It is straightforward to check that the the most general invariant polynomial can be written as a linear combination of $W^{(2)}_{3}, \, \tilde{ W}^{(2)}_{3}, \, W_2^{(2)}W_1^{(2)},$ and $\left(W_1^{(2)}\right)^3$. For convenience, these are collected in appendix~\ref{app:bigEPs}.
This procedure of using explicit codes to construct generators is carried on to $n=11$ to obtain the full set. These are given in graph form in appendix \ref{app:gengraphs}.

\paragraph{$\boldsymbol{n = 4}$}

At genus 1, we have 20 polynomials.

At genus 2, there are 45 polynomials but only 10 factorize.

9 of these polynomials derive from real codes, leaving only one fake polynomial.

\paragraph{$\boldsymbol{n = 5}$}

At genus 1, we have 395 polynomials.

At genus 2 there are 1078 polynomials, but only 23 factorize.

21 of these polynomials derive from real codes, leaving only two fake polynomials.

\paragraph{$\boldsymbol{n = 6}$}

At genus 1, we have 27,280 polynomials.

At genus 2, 79 polynomials factorize to genus 1 polynomials. We are unable to count the total number of polynomials

64 of these polynomials derive from real codes.

\paragraph{$\boldsymbol{n = 7}$}

At genus 1, we have 2,224,626 polynomials.

We are unable to count the number of genus 2 polynomials.

There are 218 polynomials from actual codes.

\subsection{Isospectral CFTs differ at genus 2}

The partition function is a coarse observable. At genus 1, it only contains information about the spectrum of the theory, and it is possible that different theories may have the same genus 1 partition function.\footnote{Such theories can also be differentiated by introducing chemical potentials for the $U(1)^n \times U(1)^n$ currents \cite{Dymarsky:2020bps}.} Higher-genus partition functions contain information about averages of OPE coefficients. In principle increasing the genus increases the amount of information extractable, though it is hoped \cite{Friedan:1986ua} that with enough partition functions, the theory will be completely specified. A more precise understanding of this idea is one of the primary motivations of the present work. 

For chiral CFTs, a demonstration is provided by Milnor's example of isospectral lattices. These correspond to chiral CFTs defined by compactification on the isospectral $d_{16}$ and $E_8^2$ lattices. It has been known since \cite{Grushevsky:2008zp} that these two theories can actually be distinguished by going to genus 5, \textit{i.e.} the partition functions are the same for $g\leqslant 4$.

In \cite{Dymarsky:2020qom} a non-chiral version of this phenomenon was discovered. This example consists of two different CFTs at $n = 7$ which have the same genus 1 partition function. These CFTs have the same spectrum, but it is clear from their definition through inequivalent $B$-form codes that they must be different. These $B$-forms are given as
\begin{equation}
    B_1=\begin{pmatrix}
    0 & 1 & 0 & 0 & 0 & 1 & 0\\
    1 & 0 & 1 & 0 & 0 & 0 & 1\\
    0 & 1 & 0 & 1 & 1 & 1 & 1\\
    0 & 0 & 1 & 0 & 1 & 0 & 0\\
    0 & 0 & 1 & 1 & 0 & 0 & 0\\
    1 & 0 & 1 & 0 & 0 & 0 & 1\\
    0 & 1 & 1 & 0 & 0 & 1 & 0
\end{pmatrix}
\, , \qquad 
B_2=\begin{pmatrix}

    0 & 1 & 0 & 0 & 0 & 1 & 1\\
    1 & 0 & 1 & 0 & 0 & 0 & 1\\
    0 & 1 & 0 & 1 & 1 & 1 & 1\\
    0 & 0 & 1 & 0 & 1 & 0 & 0\\
    0 & 0 & 1 & 1 & 0 & 0 & 0\\
    1 & 0 & 1 & 0 & 0 & 0 & 1\\
    1 & 1 & 1 & 0 & 0 & 1 & 0
\end{pmatrix}\, .
\end{equation}
It can easily be see that these matrices yield the same genus 1 enumerator polynomial,
\begin{align}
W \ =& \ y_0^7 + y_0^5 y_1^2 + 5 y_0^4 y_1^2 y_2 + 5 y_0^2 y_1^4 y_2 + y_0^5 y_2^2 + 12 y_0^3 y_1^2 y_2^2 + 9 y_0 y_1^4 y_2^2 + 4 y_0^4 y_2^3  \\ \nonumber
& + 22 y_0^2 y_1^2 y_2^3 +4 y_1^4 y_2^3 + 5 y_0^3 y_2^4 + 25 y_0 y_1^2 y_2^4 + 11 y_0^2 y_2^5 + 11 y_1^2 y_2^5 + 10 y_0 y_2^6 + 2 y_2^7\, .
\end{align}
Likewise, $W^{(2)}_1$ and $W^{(2)}_2$ can be constructed from $B_1$ and $B_2$ using \ref{eq:Wgenus2def}. We will not record them here since they are quite lengthy expressions, but they are distinct. One can further check that under factorization, both $W^{(2)}_1$ and $W^{(2)}_2$ factorize to  $(W)^2$. Holomorphic modular forms at higher genus that degenerate  in the factorization limit are referred to as cusp forms. The non-chiral analogs of such cusp forms would be $W^{(2)}_1-W^{(2)}_2$.

At $c=8$, the situation is more interesting as there are 61 pairs of isospectral theories and 5 isospectral triples. We find that the genus 2 partition functions are different for all pairs of isospectral theories. We also find that the genus 2 partition functions are different for all 5 isospectral triples. 

\section{Beyond code theories}
\label{subsec:noncode}

Throughout this paper, we have used the enumerator polynomial to simplify the form of the partition function. The primary benefit of this is that the constraints of modular invariance become very simple, allowing us to solve them exactly, \textit{i.e.}\ to write the most general partition function (in enumerator polynomial form) which satisfies the constraints. 

In our previous work \cite{Henriksson:2021qkt}, we pointed out that in fact, the partition function of every meromorphic CFT should have enumerator polynomial form, albeit with potentially negative or fractional coefficients. This follows directly from the fact that the (numerator of the) partition functions must be linear combinations of Siegel modular forms, combined with standard results relating these modular forms to code enumerator polynomials (\textit{e.g.} \cite{Runge1996}). Non-chiral CFTs are much richer however, and to our knowledge, it is not known what subset of these CFTs might have partition functions with enumerator polynomial form. Let us explore a few examples.

\subsection{Minimal models}
\label{sec:minimalmodels}
The first example is the 2D Ising CFT, which has the following genus 1 partition function:

\begin{align}
    Z^{(g = 1)}_{4,3} = \frac{1}{|\eta(\tau)|} \left( \sqrt{\theta_2(q) \theta_2(\bar q)}+ \sqrt{\theta_3(q) \theta_3(\bar q)}+ \sqrt{\theta_4(q) \theta_4(\bar q)} \right) \, .
\end{align}
The Ising CFT has $c = 1/2$, so we see that 
\begin{align}
    Z^{(g = 1)}_{4,3} = \frac{W(x_0, x_1, x_2)}{|\eta(\tau)|^{2c} } \, ,
\end{align}
where
    \begin{align}
        W(x_0, x_1, x_2) =  \sqrt{\frac{x_0 + x_1}{2}} + \sqrt{\frac{x_0 - x_1}{2}} + \sqrt{x_2}\, .
    \end{align}
The Ising model can be written as a sum of square roots of the code variables. It is easy to verify that the $S$ transformation merely cycles these terms. 

One might hope that this extends to other minimal models as well, but it appears that it does not. Consider for instance the Lee--Yang CFT, a non-unitary minimal model with $c = -22/5$. The genus 1 partition function is known,\footnote{See also \cite{Leitner:2018iyf}, which includes a calculation of the genus 2 partition function}
\begin{align}
    Z^{(g = 1)}_{5,2} = | q^{-1/60} G(q)|^2 + |q^{11/60} H(q)|^2 \, . 
\end{align}
The functions $G(q)$ and $H(q)$ are the Rogers--Ramanujan functions, defined by
\begin{align}
    G(q) = \sum_{n = 0}^\infty \frac{q^{n^2 + \frac{1}{24}}}{\eta(q)} \, , \qquad \qquad H(q) = \sum_{n = 0}^\infty \frac{q^{n^2+ n+ \frac{1}{24}}}{\eta(q)} \, .
\end{align}

This is not equal to $ \sum_{i = 2}^4 |\theta_i(q)/\eta(\tau)|^{c/2}$, as the naive pattern would suggest. It remains possible that some non-trivial identity relates this partition function to Jacobi theta functions, but we were not able to find it. As far as we can tell, the ``enumerator polynomial form'' displayed by the Ising CFT is an accident. Or perhaps it is related to the fact that the theory can be realized as a single free fermion. Perhaps the other minimal models are obtainable from a suitable generalization of New Construction A, such as the one of \cite{Yahagi:2022idq}. It would be interesting to understand this better in the future. 

\subsection{Chiral CFTs revisited}
\label{eq:extremalchiralCFTs}

As a motivating example, let us consider a simple case of meromorphic, or chiral CFTs. We use ``meromorphic'' in the sense of \cite{Schellekens:1992db}, where it is taken to mean theories where $Z = \chi(\tau) \chi(\bar \tau)$, with $\chi(\tau) = \chi(-1/\tau)$. This is a strong constraint on the form of the partition function, and it leads to the requirement that $c$ be a multiple of 8. Furthermore, if $c$ is a multiple of 24, then $\chi(\tau)$ will be individually modular invariant and therefore can be the full partition function of a chiral CFT. 

The connection between ECCs and chiral CFTs was considered in \cite{Henriksson:2021qkt}, where it was observed that every meromorphic CFT should have a partition function with enumerator polynomial form. This simply follows from the requirement that every chiral character $\chi(\tau)$ must be a modular form. For $g<4$, the ring of modular forms is completely captured by the ring of invariant polynomials after the standard substitutions $x_i \to \theta_i(q)$ \cite{Runge1996,Oura2008b}.

Let us now see some concrete examples of these observations. One interesting set of examples are the three ``extremal $c = 24$ theories.'' The first is the theory coming from the ``Golay code:''
\begin{align}
    W_{\mathrm{Golay}}(x_0, x_1) = x_0^{24} + 759 x_0^{16} x_1^8 +2576
   x_0^{12} x_1^{12}+759 x_0^8 x_1^{16} +x_1^{24}\, .
\end{align}
The Golay code is the classical binary length 24 code which maximizes the Hamming distance. Relatedly, its enumerator polynomial has the largest gap between consecutive powers of $x_0$ (\textit{i.e.} there is no $x_0^{20}$ term). The next example is the Leech lattice, with
\begin{align}
        W_{\mathrm{Leech}}(x_0, x_1) = x_0^{24}- 3  x_0^{20} x_1^4 +771  x_0^{16} x_1^8 +2558
   x_0^{12} x_1^{12}  +771  x_0^8 x_1^{16} -3 x_0^4 x_1^{20}
   +x_1^{24}\, .
\end{align}
This theory does not derive from a code through Construction A. However, it is related to the Golay code by the twisting procedure \cite{Dolan:1989kf, Dolan:1994st}, reviewed in \cite{Dymarsky:2020qom}, which relates a self-dual even lattice to a new self-dual even lattice. The Leech lattice is ``extremal'' in the sense that it is known to provide the densest possible sphere packing in 24 dimensions \cite{Cohn}. 

The final interesting example is the Monster CFT, the theory whose automorphism group is the Monster group. Its partition function is given in enumerator polynomial form by
\begin{align}
        W_{\mathrm{Monster}}(x_0, x_1) = x_0^{24}-\frac{9}{2}  x_0^{20} x_1^4 +777  x_0^{16} x_1^8 + 2549
   x_0^{12} x_1^{12} + 777 x_0^8 x_1^{16} -\frac{9}{2} x_0^4 x_1^{20}
   + x_1^{24}\, .
\end{align}
The Monster is the CFT which maximizes the spectral gap -- the gap in Virasoro primaries -- for $c = 24$. It is neither a code nor a lattice theory, but it is related to the Leech lattice by $\mathbb{Z}_2$ orbifolding. So we see that these three theories are related. 

The extremality of the Leech lattice can also be phrased in CFT language -- it is the CFT which maximizes the gap in $U(1)^c$ primaries. This remarkable connection was explored in \cite{Hartman:2019pcd}, where it was used to provide exact analytic functionals for the modular bootstrap. In fact, it turns out that the extremality of the Golay code theory also has a CFT interpretation as the maximization of the gap in $SU(2)^c$ characters. This will be explored in a future paper \cite{FutureSU2}.

We can push this further. For the Monster theory, the extremal theory (the theory with the maximium gap in Virasoro primaries) at $c = 24$, we have constructed the genus 2 and genus 3 partition function.  We can also construct the genus 1 and genus 2 partition functions of the conjectured extremal $c = 48$ theory, analogous to the Monster. We have collected a few of these lengthy expressions in appendix~\ref{app:bigEPs}.

\subsection{Non-chiral CFTs and the maximal gap}
\label{eq:extremalCFTs}

Next let us consider the more general, non-chiral case. The general strategy will be to still assume enumerator polynomial form for the partition function, but to go beyond code theories by allowing negative polynomial coefficients. The goal will be to identify interesting theories, such as the ``extremal theories'' which maximize the gap in Virasoro primaries $\Delta_{\mathrm{gap}}$. We will do this by considering first the most general (modular invariant) enumerator polynomial, and then simply choosing the coefficients $a_{i,j,k}$ to maximize the gap. We will focus on the genus 1 partition function but in a few cases with low central charge $n$, we will be able to provide genus 2 expressions as well.

\paragraph{$\boldsymbol{n = 1}$} In this case, $W_1 = x_0 + x_2$ is the only invariant polynomial. Its gap is $\Delta_{\mathrm{gap}} = 1/4$

\paragraph{$\boldsymbol{n = 2}$} Now we have a one parameter family of solutions. The gap is maximized by $W_2 = x_0^2 + x_1^2+ 2 x_2^2$, which has $\Delta_{\mathrm{gap}} = 1/2$

\paragraph{$\boldsymbol{n = 3}$} In this case, the maximal gap is $\Delta_\mathrm{gap} = 3/4$, coming from the polynomial
\begin{align}
    \tilde W_{3} = x_0^3 + 3 x_0 x_1^2 + 4 x_2^3\, .
\end{align}
We can recognize that this is the enumerator polynomial of~\eqref{eq:W3tilde}, which came from the $B$-form code corresponding to the complete graph $K_3$, the fully-connected graph with three nodes.

There is a unique genus 2 enumerator polynomial which factorizes into this genus 1, fixing the genus 2 partition function of this theory to be
\begin{align}
\begin{split}
    \tilde{W}^{(2)}_{3} &= y_0^3 + 3 y_0 y_1^2 + 3 y_0 y_2^2 + 3 y_0 y_3^2 +  6 y_1 y_2 y_3 \\
   & \quad  + 4 y_4^3 + 4 y_5^3 + 4 y_6^3 + 12 y_4 y_7^2 + 12 y_5 y_8^2 + 12 y_6 y_9^2 \, .
\end{split}
\end{align}

\paragraph{$\boldsymbol{n = 4}$} Here we find the same thing: the extremal theory, which now has $\Delta_\mathrm{gap} = 1$, also corresponds to a code theory. It has
\begin{align}
    \tilde{W}_{4} = x_0^4 + 6 x_0^2 x_1^2 + x_1^4 + 8 x_2^4 \, .
\end{align}
This theory, also realizable as the $SO(8)$ WZW with level 1, or the theory of $8$ free fermions with diagonal GSO projection, was shown to saturate the modular bootstrap constraints at $n = 4$ \cite{Collier:2016cls}. So in this case the extremal theory with enumerator polynomial form is the same as the most general extremal theory (in contrast to the cases of $n = 1$, $2$, and $3$, where the bounds of \cite{Collier:2016cls} are not saturated by our EP-form partition functions). $\tilde{W}_{4}$ is the $B$-form code deriving from the complete graph $K_4$. So in both of these cases, the extremal theories are given by the complete graphs.

Again, we can fix the genus 2 partition function from this: %
\begin{align}
\begin{split}
    \tilde{W}^{(2)}_{4} &=y_0^4 + 6 y_0^2 y_1^2 + y_{1}^4  + 6 y_0^2 y_2^2  + 6 y_0^2 y_2^2 + y_2^4  + 6 y_0^2 y_3^2 + 6 y_1^2 y_3^2  \\
    & \qquad \quad + y_3^4 + 24 y_0 y_1 y_2 y_3 + 8 y_4^4+ 8 y_5^4+ 8 y_6^4 + 8 y_7^4 \\
    & \qquad \quad + 8 y_8^4 + 8 y_9^4 + 48 y_4^2 y_7^2 + 48 y_5^2 y_8^2 + 48 y_6^2 y_9^2 \, .
\end{split}
\end{align}

\paragraph{$\boldsymbol{n = 5}$} Here something new happens. From the Molien series, the general $n = 5$ polynomial has 4 undetermined coefficients, but if we use this to cancel out the maximum number of states, the resulting ``extremal theory'' has negative degeneracies. 

Instead, we may only cancel 3, rather than 4 undetermined degeneracies. This results a function with $\Delta_{\mathrm{gap}} = 1$, and 1 remaining undetermined coefficient, 
\begin{align}
    \tilde{W}_5 = x_0^5 + \frac{16 - a_{0,0,5}}{7} x_1^4 x_2 - 4 \frac{16 - a_{0,0,5}}{7} x_0^2 x_1^2 x_2 + \ldots\, .
    \label{eq:W5}
\end{align}
If we demand that the remaining degeneracies are positive integers, we find that $a_{0,0,5}$ must be an integer satisfying $0 \leq a_{0,0,5} \leq 16$. If we set $a_{0,0,5} = 16$, we recover the code theory corresponding to the complete graph $K_5$. None of the other values of $a_{0,0,5}$ can derive from a code via New Construction A due to the signs in~\eqref{eq:W5}

\paragraph{$\boldsymbol{n = 8}$} Noting that $n = 6$ and $n = 7$ are similar to $n = 5$, we proceed to $n=8$. In this case, we find that we can cancel enough states to make $ \Delta_{\mathrm{gap}} = 7/4$, leading to a partition function with negative integer degeneracies. If we require non-negative degeneracies,\footnote{We have only checked that all Virasoro degeneracies are positive up to $\Delta = 10$, since we do not know a way to enforce it for all degeneracies in practice.} the maximum gap is $ \Delta_{\mathrm{gap}} = 5/4$. This does not saturate the bound $\Delta_\mathrm{gap} = c / 8 + 1/2$ of \cite{Collier:2016cls}.

Alternatively, we may try to only cancel scalar degeneracies. If we try to cancel the maximum number, we find $\Delta^{s = 0}_{\mathrm{gap}} = 5/2$, but then a number of other states have negative degeneracies. More interestingly, if we cancel the maximum number while requiring unitarity, we find a unique theory with $\Delta^{s = 0}_{\mathrm{gap}} = 2$. This theory is described by the enumerator polynomial
\begin{align}
    \tilde{W}^{s=0}_{8} = x_0^8 + 60 x_0^6 x_1^2+ 60 x_1^6 x_2^2 + 134 x_0^4 x_1^4-  32 x_0^4 x_2^4 - 32 x_1^4 x_2^4- 192 x_0^2 x_1^2 x_2^4 + 256 x_2^8\, .
    \label{eq:ceq8scalar}
\end{align}
Current modular bootstrap bounds require that $\Delta^{s = 0}_{\mathrm{gap}} < 2$ for $n<8$, with a kink at $\Delta^{s = 0}_{\mathrm{gap}} = 2$ for $n=8$ which is conjectured to be saturated by the $E_8$ WZW model at level 1 \cite{Collier:2016cls}. This theory holomorphically factorizes, $Z(q, \bar q) = Z_{E_8}(q)Z_{E_8}(\bar q)$ into two copies of the chiral theory which results from from compactifying 8 chiral bosons on the $E_8$ lattice. This theory is also related to the Hamming code via Construction A.

This gives another example for which a theory which is extremal in some sense can be written in EP form but does not arise from a code. That it does not derive from a code is obvious because of the negative coefficients appearing in~\eqref{eq:ceq8scalar}. This may be thought of as analogous to the enumerator-polynomial form representation of the Leech lattice and Monster partition functions.

A third interesting example exists for $n = 8$ -- the theory described in \cite{Bae:2017kcl} with $O^+_{10}(2).2$ automorphism group. This theory is completely fixed from the general invariant EP form by requiring that our partition functions have only integer scaling dimensions, and that the degeneracy of scalars at $\Delta = 1$ is 496. The resulting enumerator polynomial is 
\begin{align}
    W^{O^+_{10}(2).2}_{n = 8} = x_0^8 - 2 x_0^6 x_1^2 + 10 x_0^4 x_1^4 - 2 x_0^2 x_1^6 + x_1^8 + 30 x_0^4 x_2^4 + 180 x_0^2 x_1^2 x_2^4 + 30 x_1^4 x_2^4 + 8 x_2^8 \, .
\end{align}

In principle, it should be possible to check if these genus 1 partition functions correspond to a single genus 2 partition function. Since these theories do not derive from codes, it may be that their genus 1 partition functions have EP form but their genus 2 partition functions do not -- we do not know if this is possible. For the chiral CFTs, we found that extremal genus 2 partition functions at $c = 24$ is entirely fixed by factorization,
so it would be interesting to try that here. However, the most general genus 2 polynomial is beyond what we are able to do with our desktop computers right now, so we will have to leave exploring them to the future.

\paragraph{$\boldsymbol{n = 24}$} In this case, by allowing for negative degeneracies we can construct a modular invariant function with $\Delta_{\mathrm{gap}} = 5$. As an interesting side note, it is possible to identify the theory $Z = Z_\mathrm{Monster}(q)Z_\mathrm{Monster}(\bar q)$ by setting quarter integer degeneracies to 0 and requiring a gap. However, this modular-invariant partition function has negative degeneracies from the point of view of a non-chiral CFT because the $1-q$ in the vacuum character of $Z_\mathrm{Monster}(q)$ has a minus sign, which multiplies all of the characters in $Z_\mathrm{Monster}(\bar q)$, and vice-versa.

\subsection[Asymptotics at large $c$]{Asymptotics at large $\boldsymbol{c}$}

Finally, let us make a few comments about theories in the limit of large central charge.

\subsubsection{Complete graphs}

The complete graphs provide a special class of code theories, which includes every positive integer $n$. We can study them by explicitly constructing their enumerator polynomials from their $B$-matrices, which have a 1 in every entry except on the diagonal. For $c \geqslant 4$, the gap of very complete-graph theory has $\Delta_{\textrm{gap}} = 1$. In fact, this is the upper bound on the gap of any code theory \cite{Dymarsky:2020qom}. 

As we noted before, the $n = 4$ deriving from the complete graph $K_4$ is equivalent to the $SO(8)$ WZW model. It appears that this pattern continues beyond $n =4$. We have checked for $n$ up to 15 that the number of currents is always given by $N_n = n(2n-1)$. Extra currents require extra symmetries, and $n(2n-1)$ is the number of currents one would find in the presence of symmetry enhancement to $SO(2n)$. In fact, for large enough $n$, no other rank $n$ Lie group contains this many currents, so $SO(2n)$ is the only possibility. As a result, we conjecture that the complete graph codes give a family of CFTs with symmetry enhancement to $SO(2n)$. It would be interesting to study these theories further, to try to prove this conjecture or understand their holographic duals, in the future. 

\subsubsection{Estimate for non-unitarity theories}

It is also possible to provide an estimate of the maximal gap possible by comparing the number of degeneracies we need to cancel with the number of coefficients we can choose. That is, if we have $k$ coefficients, then we should be able to set $k$ degeneracies to zero.\footnote{This is not rigorous because in principle it is possible that the coefficients do not appear in the degeneracies independently, which could prevent us from canceling some degeneracies. In practice we see that this does not happen, at least up to $n = 24$. Still, one should take this subsection as a speculation rather than a proof.}

First we consider the number of possible operators up to a given scaling dimension. Because these theories have quarter-integer scaling dimensions, the maximal number of possible operators up to $\Delta$ given by 
\begin{align}
        \rho(\Delta) = 4 \sum_{k=0}^\Delta (k + 1) = 2 (\Delta + 2)(\Delta+1) \, .
\end{align}
So to create a gap $\Delta$, one must cancel $2 (\Delta + 2)(\Delta+1)$ theories.

Next we determine the number of coefficients which we can pick to cancel degeneracies. This is given by the coefficients appearing in the Molien series. The $n^{\text{th}}$ such coefficient is given by
\begin{align}
        \omega(n) = \frac{1}{72} \left( 47 + 9 (-1)^n + 6n(6+n) + 16 \cos \frac{2 n \pi}{3} \right)
        \, .
\end{align}
At large $n$, $\omega(n) \sim n^2 / 12$ and $\rho(\Delta) \sim 2 \Delta^2$. Thus we can create a gap up to $\Delta$ at central charge $c = n$ as long as $\omega(n) = \rho(\Delta)$, leading to the estimate
\begin{align}
        \Delta_{\textrm{gap}} = \frac{n}{\sqrt{24}} \, .
\end{align}
This is a huge gap, and is inconsistent with the strongest current asymptotic bounds given in \cite{Afkhami-Jeddi:2019zci}. However, those bounds require unitarity. So we expect that if such large gaps are possible then they must come from modular invariant functions with negative degeneracies of Virasoro characters. This is exactly what we find for $n = 24$. In that case, $\Delta_{\mathrm{gap}} = 5$ but the function with that gap has negative degeneracies. Thus our estimate on gap avoids the bootstrap bounds because the modular invariant functions satisfying it will be non-unitarity.

\section{Discussion}

This paper is an invitation to study the consequences of higher-genus modular invariance, using the example of theories defined from quantum error-correcting codes through the New Construction A of \cite{Dymarsky:2020qom}. The guiding idea is that higher-genus modular invariance should be more constraining than genus 1. Error-correcting code theories provide a simple playground to explore this because their partition functions conform to the polynomial ansatz central to this paper.
In this simple ``enumerator-polynomial form,'' the action of the modular group $SP(2g, \mathbb{Z})$ becomes a set of simple linear transformations on the polynomial variables. This allows for a simple algorithm to solve the constraints and list every partition function which could possibly derive from a code. This also requires that we impose (1) that all polynomial coefficients are positive integers, as they should derive from degeneracies of codewords, and (2) that the higher-genus partition functions factorize into lower-genus partition functions in the limit where Riemann surface they live on becomes singular.

We find that higher-genus modular invariance is very constraining. Specifically, there are a large number of genus 1 expressions which look like consistent partition functions. We have proven that such partition functions \emph{cannot arise, via factorization, from genus 2 partition functions with enumerator polynomial form}. This means that they cannot be code theories, but, importantly, it does not imply that they cannot be CFTs at all. Since the number of actual $B$-form codes is known, we can compare our results to the true set of code theories. We find that for $n = 3$, genus 2 modular invariance rules out all polynomials which do not derive from codes. For $n>3$, the number of fake polynomials is greatly reduced by genus 2 considerations, and presumably pushing our algorithm to higher genus would be required to remove all fake polynomials. It would be interesting to understand exactly what genus is required for a given $n$.\footnote{For chiral theories, it is known that the required genus is at most $n / 2 - 1$. See \cite{Runge1996, Henriksson:2021qkt}.}

In \cite{Dymarsky:2020qom}, a number of isospectral code CFTs were discovered -- theories which are known to be different, but which have the same genus 1 partition function and thus the same spectrum. Such theories are non-chiral analogs of Milnor's example isospectral Euclidean lattices in 16 dimensions, which define isospectral chiral CFTs. We find that all of the examples given in \cite{Dymarsky:2020qom} of isospectral theories up to $n = 8$ can be distinguished by their genus 2 partition functions. This is similar to how Milnor's example can be distinguished by their genus 5 partition functions \cite{Grushevsky:2008zp}.

One of the main lessons of this paper is that the enumerator polynomial ansatz for the partition function greatly simplifies a number of questions. One direction we have investigated is the maximum gap of theories with enumerator polynomial form. Theories deriving from actual codes have a maximal gap of $1$. In the case of $n = 4$, $\Delta_\mathrm{gap} =1$ is the actual upper bound on the gap \cite{Collier:2016cls}, and the theory saturating it, the $SO(8)$ WZW model, is known to be a code theory \cite{Dymarsky:2020qom}. However it is possible that there are non-code theories with enumerator polynomial form, which occurs for chiral CFTs, for instance, in the case of the Monster CFT. For $n>4$, we found that a larger gap is consistent with enumerator polynomial form. If we allow for negative Virasoro degeneracies, we can construct very large gaps (ex. $\Delta_\mathrm{gap} = 5$ for $n = 24$). For the more interesting case of unitary theories, we were not able to say as much. This is because we do not have a simple way to impose positivity over all Virasoro primaries, as they are not related to the code variables in a simple way. If such a method could be found, we believe it could be very interesting to investigate the maximum gap of theories with enumerator polynomial form.

We found a few other examples of known theories with enumerator polynomial form. Two $n = 8$ theories saturating various bootstrap bounds were shown to be non-code theories whose partition functions have enumerator polynomial form. It is notable that such theories only appear at $n = 4$ and $n = 8$. There are known theories saturating the bounds of \cite{Collier:2016cls, Bae:2017kcl} at other values of $c$ -- in particular, these include a number of other WZW models. These cannot have enumerator polynomial form because their scaling dimensions are not all quarter-integers. It would be very interesting to try to see if such theories might derive from some generalization of the New Construction A.
An important step in this direction was taken in \cite{Yahagi:2022idq}, where a construction was given for error-correcting codes over fields $F_p$ for any prime $p$. See also \cite{FutureAnatoly}.

In principle, it should be possible to extract information about OPE coefficients from the higher-genus partition function -- see \cite{Keller:2017iql} for an example where this is done for chiral theories. We think it would be very interesting to try to do this for the theories considered in this paper. This could lead to a better understanding of the relation between higher-genus constraints and the constraints coming from the standard conformal bootstrap of sphere four-point functions. Ultimately, the goal of pursuing this direction would be to understand if the higher-genus partition functions give enough information to fully specify the CFT, in the spirit of \cite{Friedan:1986ua}.

Another potential direction would be to study the relation between the partition functions of code theories, and the set of modular forms. In the chiral case, where we consider only holomorphic modular forms, it is known that enumerator polynomials, with Jacobi-thetas for arguments, give the full set of modular forms for $g < 5$. To our knowledge, the relationship for non-chiral CFTs and modular functions of two variables is completely unknown. As a result, we do not know anything about the relation between the set of enumerator polynomial theories, and the full set of 2d CFTs. Understanding this further could be a useful step in classifying 2d CFTs. 

Finally, it would be interesting to use our work to address quantum gravity. AdS${}_3$/CFT${}_2$ is one of the most useful testing grounds for new ideas about holography. Recently, this has included new insights about the significance of averaging over the moduli space of CFTs to define a particular bulk theory \cite{Afkhami-Jeddi:2020ezh, Maloney:2020nni}. In a future work \cite{FutureSU2} we will apply some of these insights to the case of chiral CFTs.
In \cite{Dymarsky:2020qom}, a formula for the average over $B$-form codes was given, and the holographic interpretation was discussed in \cite{Dymarsky:2020pzc}. It would be interesting to try to discover the genus 2 average for $B$-form codes, and understand its holographic implications.

\section*{Acknowledgments} 

We thank Anatoly Dymarsky for many useful conversations and for comments on an earlier draft of this paper. This project has received funding from the European Research Council (ERC) under the European Union's Horizon 2020 research and innovation programme (grant agreement no.~758903).

\appendix


\begin{figure}[h]
\centering
\raisebox{-4pt}{\includegraphics[width=0.2\textwidth]{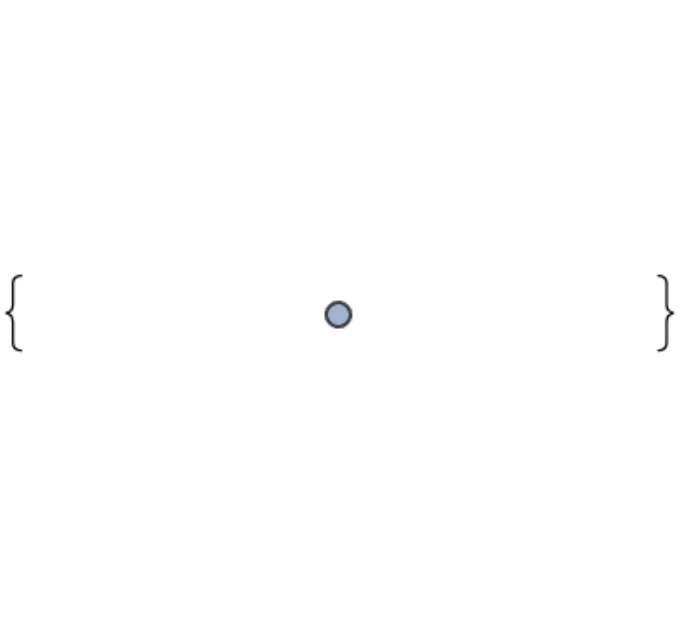}}\hspace{24pt} \raisebox{31pt}{\includegraphics[width=0.2\textwidth]{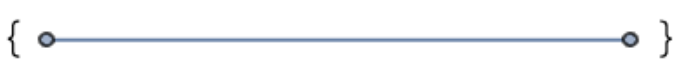}}\hspace{24pt}\includegraphics[width=0.4\textwidth]{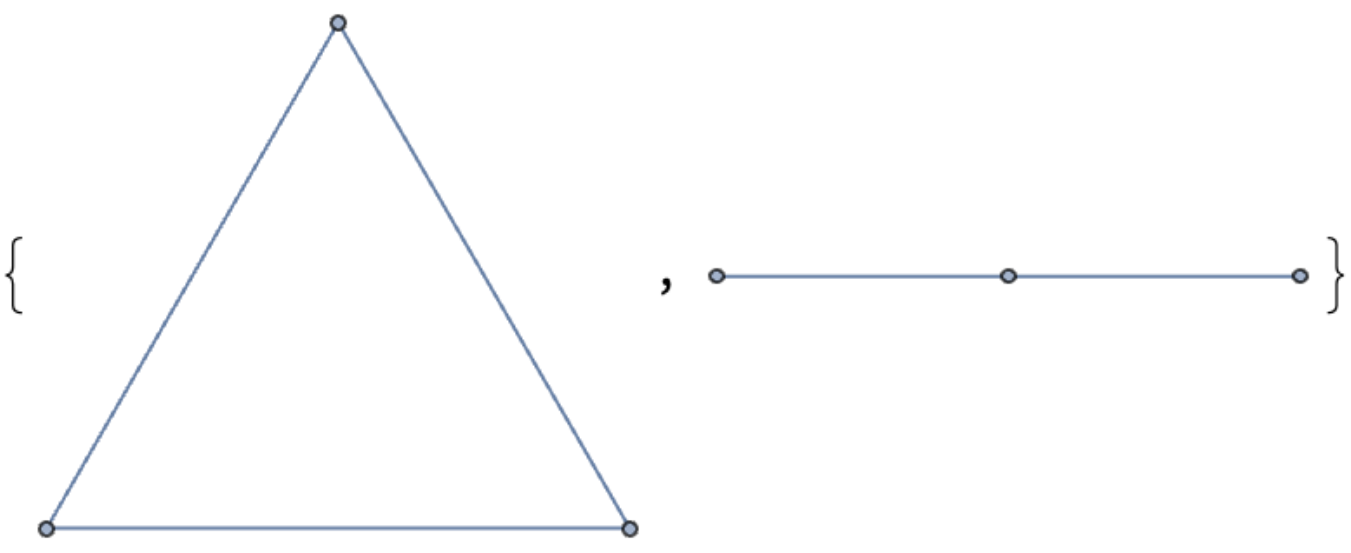}
\caption{$Q^{[2]}_1$ (left), $Q^{[2]}_2$ (middle), and $Q^{[2]}_3$ (right).}
\label{fig:gen1-3}
\end{figure}



\begin{figure}[h]
\centering
\includegraphics[width=0.6\textwidth]{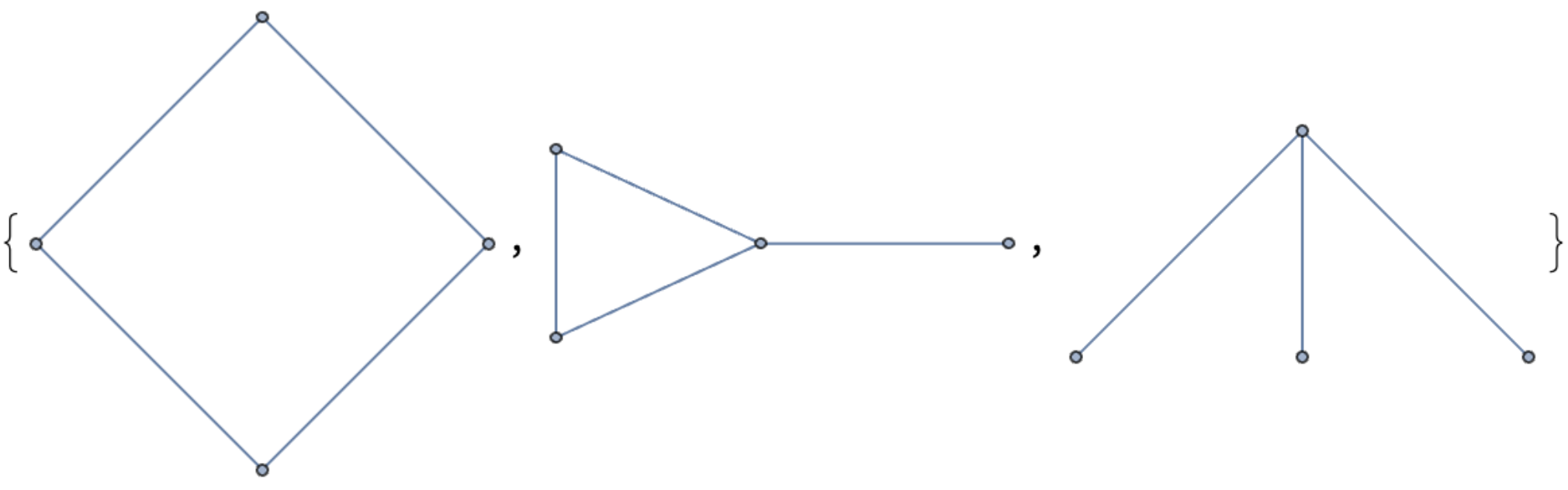}
\caption{$Q^{[2]}_4$  }
\end{figure}

\begin{figure}[h]
\centering
\includegraphics[width=0.8\textwidth]{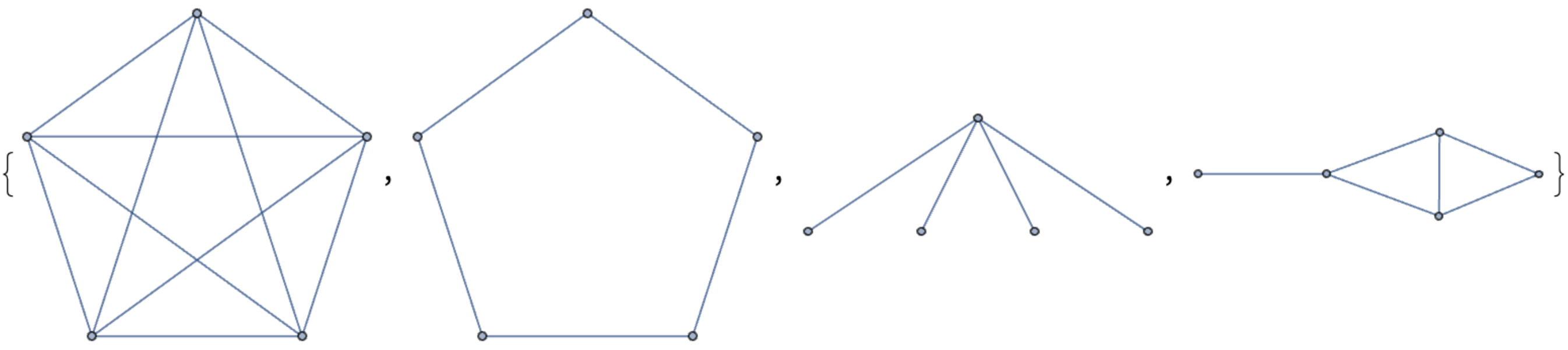}
\caption{$Q^{[2]}_5$  }
\end{figure}

\begin{figure}[h]
\centering
\includegraphics[width=0.8\textwidth]{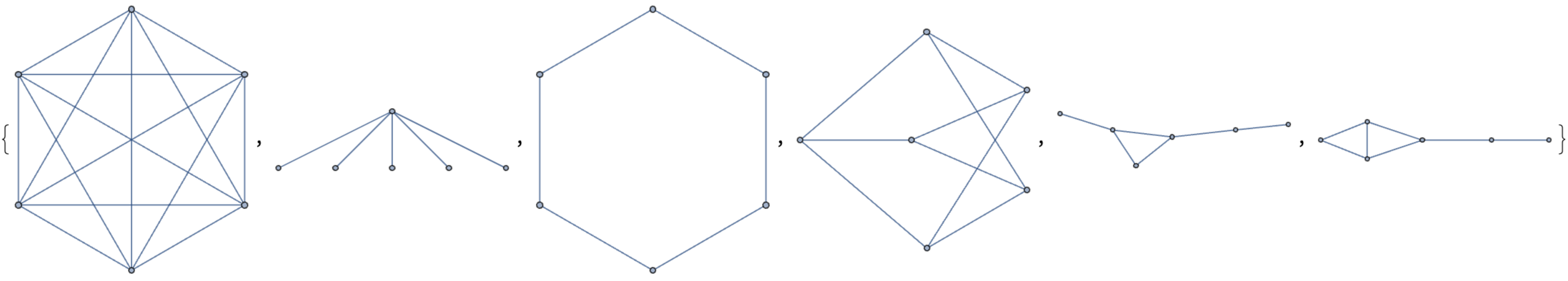}
\caption{$Q^{[2]}_6$  }
\end{figure}

\begin{figure}[h]
\centering
\includegraphics[width=0.8\textwidth]{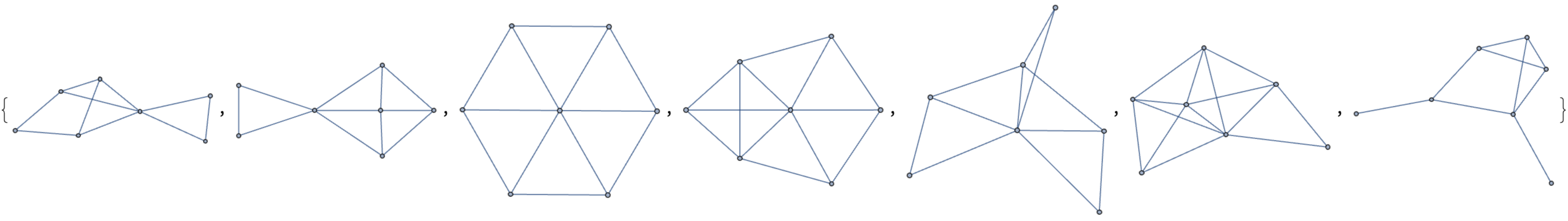}
\caption{ $Q^{[2]}_7$ }
\end{figure}

\begin{figure}[h]
\centering
\includegraphics[width=0.8\textwidth]{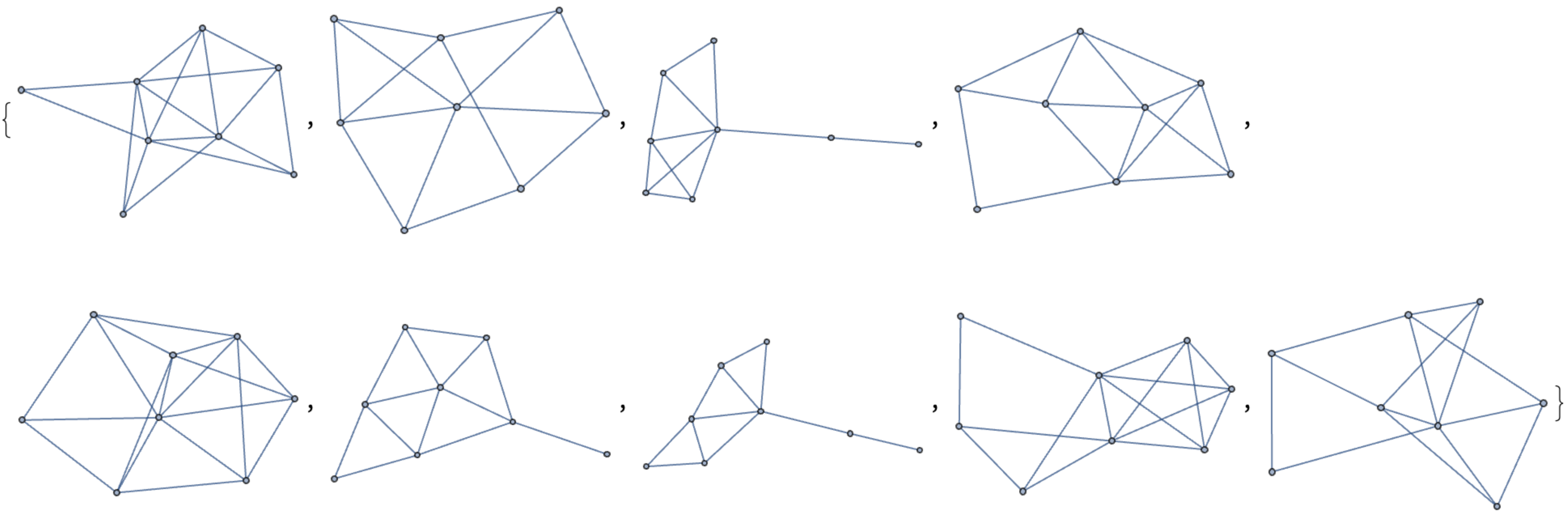}
\caption{$Q^{[2]}_8$ }
\label{fig:gen8}
\end{figure}

\begin{figure}[h]
\centering
\includegraphics[width=0.8\textwidth]{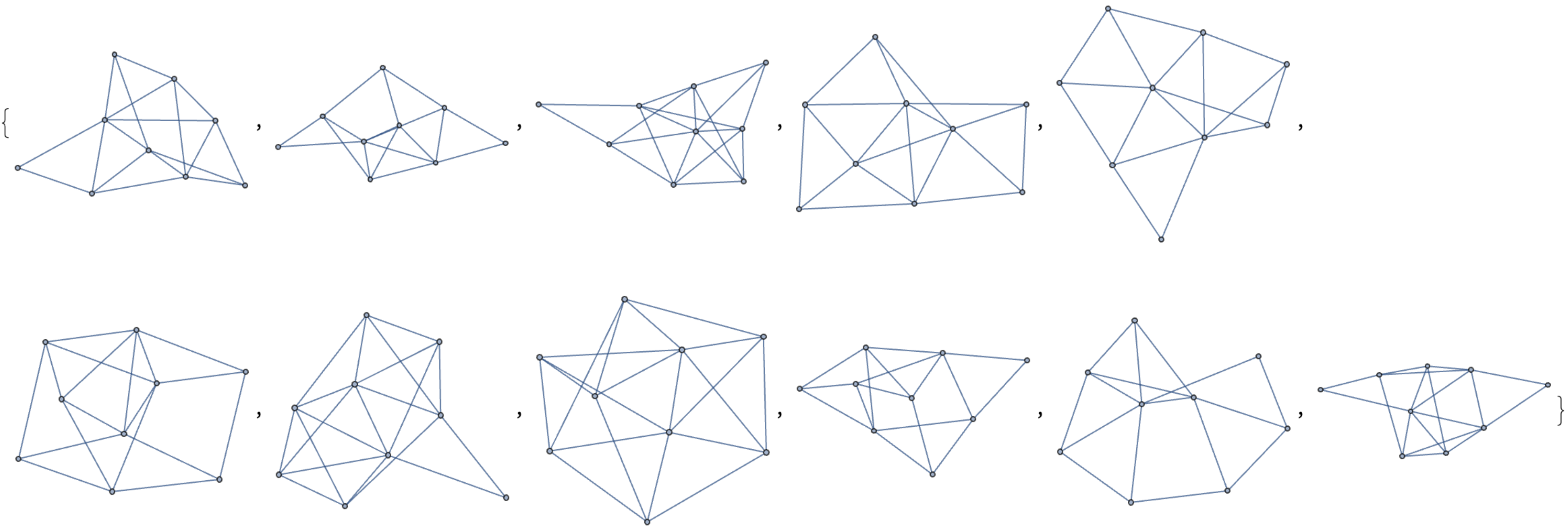}
\caption{$Q^{[2]}_9$ }
\label{fig:gen9}
\end{figure}
\begin{figure}[h]
\centering
\includegraphics[width=0.8\textwidth]{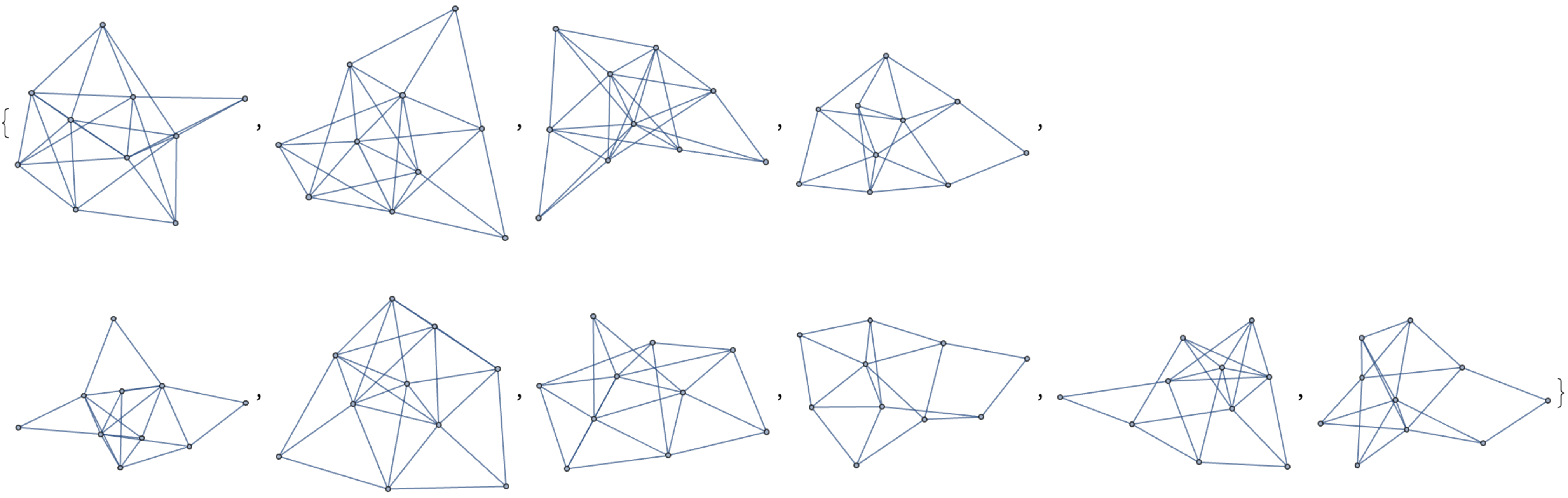}
\caption{$Q^{[2]}_{10}$ }
\label{fig:gen10}
\end{figure}
\begin{figure}[h]
\centering
\includegraphics[width=0.8\textwidth]{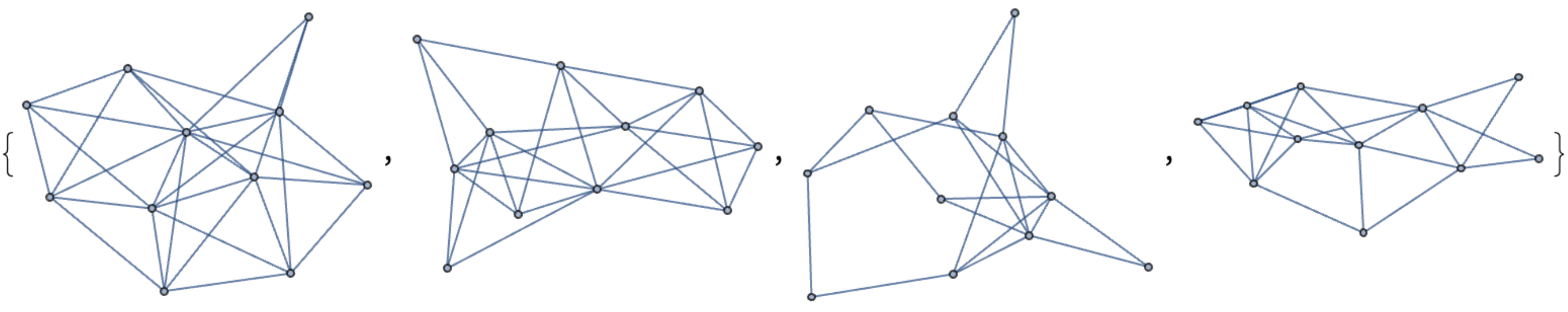}
\caption{$Q^{[2]}_{11}$ }
\label{fig:gen11}
\end{figure}

\section{Generators for invariant ring at genus 2}
\label{app:gengraphs}

In section~\ref{sec:results} we analyzed the generators for ring $R_g$ of invariant genus 2 polynomials. It is convenient to organize the generators by degree. Consider the set $Q_n^{[2]}$ of generators of degree $n$, then the ring is given by
\begin{equation}
    R_2=\left\langle Q_1^{[2]}\cup Q_2^{[2]}\cup \cdots \cup Q_{11}^{[2]}\right\rangle
\,.
\end{equation}
We represent each generator by an undirected graph, and show the generators in figures \ref{fig:gen1-3}--\ref{fig:gen11}.

\section{Genus 2 partition functions in enumerator polynomial form}
\label{app:bigEPs}

Here we include a few expressions too big to conveniently include in the main text.

\subsection{$n = 3$ theories}

First let us provide the genus 2 partition functions of each of the actual $n = 3$
code theories. 
\begin{align}
\begin{split}
    W_1^3 &= (x_0 + x_2)^3 \\
    \left(W_1^{(2)}\right)^3 &= y_0^3+3 y_4 y_0^2+3 y_5 y_0^2+3 y_6 y_0^2+3 y_4^2 y_0+3 y_5^2 y_0+3 y_6^2 y_0 \\
    & \qquad +6 y_4 y_5 y_0+6 y_4 y_6 y_0+6 y_5 y_6 y_0+y_4^3+y_5^3+y_6^3+3 y_4 y_5^2 \\
    & \qquad +3 y_4 y_6^2+3 y_5 y_6^2+3
   y_4^2 y_5+3 y_4^2 y_6+3 y_5^2 y_6+6 y_4 y_5 y_6
\end{split}
\end{align}

\begin{align}
\begin{split}
    W_1 W_2 &= (x_0 + x_2)(x_0^2 + x_1^2 + 2 x_2^2) \\
    W_1^{(2)} W_2^{(2)} &= y_0^3+y_4 y_0^2+y_5 y_0^2+y_6 y_0^2+y_1^2 y_0+y_2^2
   y_0+y_3^2 y_0+2 y_4^2 y_0+2 y_5^2 y_0 \\
   & \qquad +2 y_6^2
   y_0+2 y_7^2 y_0+2 y_8^2 y_0+2 y_9^2 y_0+2 y_4^3+2
   y_5^3+2 y_6^3+2 y_4 y_5^2 \\
   & \qquad +2 y_4 y_6^2+2 y_5
   y_6^2+2 y_4 y_7^2+2 y_5 y_7^2+2 y_6 y_7^2+2 y_4
   y_8^2+2 y_5 y_8^2+2 y_6 y_8^2 \\
   & \qquad +2 y_4 y_9^2+2 y_5
   y_9^2+2 y_6 y_9^2+y_1^2 y_4+y_2^2 y_4+y_3^2
   y_4+y_1^2 y_5+y_2^2 y_5+y_3^2 y_5 \\
   & \qquad +2 y_4^2
   y_5+y_1^2 y_6+y_2^2 y_6+y_3^2 y_6+2 y_4^2 y_6+2
   y_5^2 y_6
\end{split}
\end{align}

\begin{align}
\begin{split}
    W_3 &= x_0^3 + 3 x_1^2 x_2 + 3 x_0 x_2^2 + x_2^3 \\
    W_3^{(2)} &= y_0^3+3 y_4^2 y_0+3 y_5^2 y_0+3 y_6^2
   y_0+y_4^3+y_5^3+y_6^3+3 y_5 y_7^2+3 y_6 y_7^2+3
   y_4 y_8^2  \\
   & \qquad +3 y_6 y_8^2 +3 y_4 y_9^2+3 y_5 y_9^2+3
   y_1^2 y_4+3 y_2^2 y_5+3 y_3^2 y_6 \\
   & \qquad +6 y_4 y_5 y_6+6
   y_3 y_7 y_8+6 y_2 y_7 y_9+6 y_1 y_8 y_9
\end{split}
\end{align}

\begin{align}
\begin{split}
    \tilde{W}_3 &= x_0^3 + 3 x_0 x_1^2 + 4 x_2^3 \\
    \tilde{W}_3^{(2)} &= y_0^3+3 y_1^2 y_0+3 y_2^2 y_0+3 y_3^2 y_0+4 y_4^3+4
   y_5^3+4 y_6^3+12 y_4 y_7^2\\ 
   & \qquad +12 y_5 y_8^2+12 y_6
   y_9^2+6 y_1 y_2 y_3
\end{split}
\end{align}

\subsection{Chiral theories}
The genus 1 partition function for the extremal $n = 24$ chiral theory, also known as the Monster CFT, is 
\begin{align}
\begin{split}
    \tilde W^{(1)}_{24}(x_0, x_1) \ &= \ x_0^{24}-\frac{9}{2} x_1^4 x_0^{20}+777 x_1^8 x_0^{16}+2549
   x_1^{12} x_0^{12}+777 x_1^{16} x_0^8-\frac{9}{2} x_1^{20}
   x_0^4+x_1^{24}
\end{split}
\end{align}
while its genus 2 partition function is 
\begin{align}
\begin{split}
    \tilde W^{(2)}_{24}&(x_0, x_1,x_2, x_3) \ = \ \frac{1}{4} \Big(4 x_0^{24}-18 \left(x_1^4+x_2^4+x_3^4\right)
   x_0^{20}+63 x_1^2 x_2^2 x_3^2 x_0^{18}
   \\
   &+12 \left(259 x_1^8+3
   \left(x_2^4+x_3^4\right) x_1^4+259 x_2^8+259 x_3^8+3 x_2^4
   x_3^4\right) x_0^{16}-9396 x_1^2 x_2^2 x_3^2
   \left(x_1^4+x_2^4+x_3^4\right) x_0^{14}\\
   & +2 \big(5098
   x_1^{12} -2313 \left(x_2^4+x_3^4\right) x_1^8-3 \left(771
   x_2^8-152918 x_3^4 x_2^4+771 x_3^8\right) x_1^4+5098
   x_2^{12}+5098 x_3^{12}\\
   & -2313 x_2^4 x_3^8-2313 x_2^8
   x_3^4\big) x_0^{12}-6 x_1^2 x_2^2 x_3^2 \big(6105
   x_1^8-233438 \left(x_2^4+x_3^4\right) x_1^4+6105 x_2^8+6105
   x_3^8 \\
   & -233438 x_2^4 x_3^4\big) x_0^{10}+6 \big(518
   x_1^{16}-771 \left(x_2^4+x_3^4\right) x_1^{12}+6 \left(2863
   x_2^8+142508 x_3^4 x_2^4+2863 x_3^8\right) x_1^8 \\
   & +\left(-771
   x_2^{12}+855048 x_3^4 x_2^8+855048 x_3^8 x_2^4-771
   x_3^{12}\right) x_1^4+518 x_2^{16}+518 x_3^{16}-771 x_2^4
   x_3^{12}\\
   &+17178 x_2^8 x_3^8-771 x_2^{12} x_3^4\big)
   x_0^8-12 x_1^2 x_2^2 x_3^2 \big(783 x_1^{12}-116719
   \left(x_2^4+x_3^4\right) x_1^8-\big(116719 x_2^8 \\
   & +1315158
   x_3^4 x_2^4+116719 x_3^8\big) x_1^4+783 x_2^{12}+783
   x_3^{12}-116719 x_2^4 x_3^8-116719 x_2^8 x_3^4\big)
   x_0^6-6 \big(3 x_1^{20}\\ 
   &-6 \left(x_2^4+x_3^4\right)
   x_1^{16}+\left(771 x_2^8-152918 x_3^4 x_2^4+771 x_3^8\right)
   x_1^{12}+\big(771 x_2^{12}-855048 x_3^4 x_2^8-855048 x_3^8
   x_2^4\\ 
   & +771 x_3^{12}\big)  x_1^8-2 \left(3 x_2^{16}+76459
   x_3^4 x_2^{12}+427524 x_3^8 x_2^8+76459 x_3^{12} x_2^4+3
   x_3^{16}\right) x_1^4+3 \big(x_2^{20}-2 x_3^4 x_2^{16} \\
   & +257
   x_3^8 x_2^{12}+257 x_3^{12} x_2^8-2 x_3^{16}
   x_2^4+x_3^{20}\big)\big) x_0^4+3 x_1^2 x_2^2 x_3^2
   \big(21 x_1^{16}-3132 \left(x_2^4+x_3^4\right) x_1^{12}-2
   \big(6105 x_2^8 \\
   & -233438 x_3^4 x_2^4+6105 x_3^8\big)
   x_1^8-4 \left(783 x_2^{12}-116719 x_3^4 x_2^8-116719 x_3^8
   x_2^4+783 x_3^{12}\right) x_1^4+3 \big(7 x_2^{16}\\
   & -1044
   x_3^4 x_2^{12}-4070 x_3^8 x_2^8-1044 x_3^{12} x_2^4+7
   x_3^{16}\big)\big) x_0^2+2 \big(2 x_1^{24}-9
   \left(x_2^4+x_3^4\right) x_1^{20}+6 \big(259 x_2^8\\
   & +3 x_3^4
   x_2^4+259 x_3^8\big) x_1^{16}+\left(5098 x_2^{12}-2313
   x_3^4 x_2^8-2313 x_3^8 x_2^4+5098 x_3^{12}\right) x_1^{12}+3
   \big(518 x_2^{16}\\
   & -771 x_3^4 x_2^{12}+17178 x_3^8 x_2^8-771
   x_3^{12} x_2^4+518 x_3^{16}\big) x_1^8-9 \big(x_2^{20}-2
   x_3^4 x_2^{16} +257 x_3^8 x_2^{12}+257 x_3^{12} x_2^8\\
   & -2
   x_3^{16} x_2^4+x_3^{20}\big) x_1^4+2 x_2^{24}+2 x_3^{24}-9
   x_2^4 x_3^{20}+1554 x_2^8 x_3^{16}+5098 x_2^{12}
   x_3^{12}+1554 x_2^{16} x_3^8-9 x_2^{20} x_3^4\big)\Big)
\end{split}
\end{align}

The $c = 48$ extremal theory has partition function in EP form given by
\begin{align}
\begin{split}
    \tilde W^{(1)}_{48}&(x_0, x_1) \ = \ x_0^{48}-9 x_1^4 x_0^{44}+\frac{1155}{32} x_1^8
   x_0^{40}+\frac{41641}{4} x_1^{12} x_0^{36}+\frac{4314189}{8}
   x_1^{16} x_0^{32}\\
   & \qquad +\frac{16161123}{4} x_1^{20}
   x_0^{28}+\frac{121555689}{16} x_1^{24}
   x_0^{24}+\frac{16161123}{4} x_1^{28}
   x_0^{20}\\ 
   & \qquad +\frac{4314189}{8} x_1^{32} x_0^{16}
   +\frac{41641}{4}
   x_1^{36} x_0^{12}+\frac{1155}{32} x_1^{40} x_0^8-9 x_1^{44}
   x_0^4+x_1^{48}
\end{split}
\end{align}

We have also obtained the polynomial expression for the genus 3 partition function of the $c = 24$ extremal theory, and for the genus 2 partition function of the conjectured $c = 48$ extremal theory. These were too large to included even here but are available upon request.

\bibliography{cite.bib}

\bibliographystyle{JHEP.bst}

\end{document}